\documentclass[fleqn,10pt]{wlscirep}
\usepackage[utf8]{inputenc}
\usepackage[T1]{fontenc}
\usepackage{bm}
\usepackage{siunitx}
\usepackage{lettrine}
\usepackage{makecell} 
\DeclareUnicodeCharacter{2212}{-}
\usepackage[utf8]{inputenc}
\usepackage[nohyperlinks, nolist]{acronym}
\usepackage{hyperref}
\usepackage{graphicx}
\usepackage{amsmath}
\usepackage{amssymb}
\usepackage{verbatim}
\usepackage[]{algorithm2e}
\usepackage{subcaption}
\usepackage{float}
\usepackage{multirow}
\usepackage{multicol}

\newcommand{\orcid}[2]{\href{https://orcid.org/#2}{ \textcolor{orcid}{#1}}}

\newcommand{\NSTARS}{820}%
\newcommand{\NCADENCES}{1004}%
\newcommand{\NBACKGROUND}{14,711}%

\newcommand{\NTOPCAND}{8}

\definecolor{dcol}{rgb}{0.7, 0.1, 0.1}
\definecolor{orcid}{rgb}{0.06, 0.4, 1.0}

\title{A deep-learning search for technosignatures of $\NSTARS$ nearby stars}

\author[1,2,3*]{\orcid{Peter Xiangyuan  Ma}{0000-0001-8975-3719}}
\author[3,4,5]{\orcid{Cherry Ng}{0000-0002-3616-5160}}
\author[6]{\orcid{Leandro Rizk}{0000-0003-4573-3534}}
\author[4,5]{\orcid{Steve Croft}{0000-0003-4823-129X}}
\author[4,5,7,8]{\orcid{Andrew P. V. Siemion}{0000-0003-2828-7720}}
\author[4]{\orcid{Bryan Brzycki}{0000-0002-7461-107X}}
\author[4]{\orcid{Daniel Czech}{0000-0002-8071-6011}}
\author[9]{Jamie Drew}
\author[4]{\orcid{Vishal Gajjar}{0000-0002-8604-106X}}
\author[4]{\orcid{John Hoang}{0000-0001-5591-5927}}
\author[4, 10]{\orcid{Howard Isaacson}{0000-0002-0531-1073}}
\author[4]{\orcid{Matt Lebofsky}{0000-0002-7042-7566}}
\author[4]{David MacMahon}
\author[4]{\orcid{Imke de Pater}{0000-0002-4278-3168}}
\author[11,4]{\orcid{Danny C. Price}{0000-0003-2783-1608}}
\author[4]{\orcid{Sofia Z. Sheikh}{0000-0001-7057-4999}}
\author[9]{S. Pete Worden}

\affil[1]{Department of Mathematics, University of Toronto, 40 St. George Street, Toronto, ON M5S 2E4, Canada}

\affil[2]{Department of Physics, University of Toronto, 60 St. George Street, Toronto, ON M5S 1A7, Canada}
\affil[3]{Dunlap Institute for Astronomy \& Astrophysics, University of Toronto, 50 St.~George Street, Toronto, ON M5S 3H4, Canada}

\affil[4]{Radio Astronomy Laboratory, 501 Campbell Hall, University of California,
Berkeley, CA 94720, USA}

\affil[5]{SETI Institute, Mountain View, CA 94043, USA}

\affil[6]{David A.~Dunlap Department of Astronomy \& Astrophysics, University of Toronto, 50 St. George Street, Toronto, ON M5S 3H4, Canada}

\affil[7]{Jodrell Bank Centre for Astrophysics (JBCA), Department of Physics \& Astronomy, Alan Turing Building, The University of Manchester, M13 9PL, UK}

\affil[8]{University of Malta, Institute of Space Sciences and Astronomy}
\affil[9]{Breakthrough Initiatives, Moffett Field, CA 94035, USA}
\affil[10]{Centre for Astrophysics, University of Southern Queensland, Toowoomba, QLD, Australia}

\affil[11]{International Centre for Radio Astronomy Research, Curtin University, Bentley WA 6102, Australia}

\affil[*]{peterxy.ma@mail.utoronto.ca}

\begin{abstract}
The goal of the Search for Extraterrestrial Intelligence (SETI) is to quantify the prevalence of technological life beyond Earth via their ``technosignatures". One theorized technosignature \textcolor{black}{is} narrowband Doppler drifting radio signals. The principal challenge in conducting SETI in the radio domain is developing a generalized technique to reject human radio frequency interference (RFI). Here, we present the most comprehensive deep-learning based technosignature search to date, returning $\NTOPCAND$ promising ETI \textcolor{black}{signals of interest} for re-observation as part of the Breakthrough Listen initiative. The search comprises $\NSTARS$ unique targets observed with the Robert C. Byrd Green Bank Telescope, totaling over 480\,hr of on-sky data. We implement a novel $\beta-$Convolutional Variational Autoencoder to \textcolor{black}{identify} technosignature candidates in a semi-unsupervised manner \textcolor{black}{while keeping the false positive rate manageably low}.
This new approach presents itself as a leading solution in accelerating SETI and other transient research into the age of data-driven astronomy.

\end{abstract}
\begin{document}
\flushbottom

\maketitle
\date{
\textcolor{red}{Submission date: December 10, 2021 Nature Astronomy, Accepted November 30, 2022}
}
\thispagestyle{empty}

\lettrine{``A}{re}  we alone?'' is one of the most profound scientific questions humans have asked. The search for extraterrestrial intelligence (SETI) aims to answer this question by looking for evidence of intelligent life elsewhere in the galaxy via the ``technosignatures'' created by their technology. The majority of technosignature searches to date have been conducted at radio frequencies, given the ease of propagation of radio signals through interstellar space \cite{Cocconi1959}, as well as the relative efficiency of construction of powerful radio transmitters and receivers. 
\textcolor{black}{One type of} technosignatures that is most readily distinguishable from natural astrophysical radio emissions \textcolor{black}{is that of} narrow-band (on the order of 1\,Hz) and / or exhibit Doppler drifts due to the relative motions of transmitter and receiver \cite{Tarter2001}.
The detection of an unambiguous technosignature would demonstrate the existence of ETI and is thus of acute interest both to scientists and the general public.

Currently, one of the main driving forces of SETI research is the Breakthrough Listen (BL) Initiative\footnote{https://breakthroughinitiatives.org}. Since 2016, BL has been using the Robert C.\ Byrd Green Bank Telescope (GBT) in the U.S. and the Parkes ``Murriyang'' Telescope in Australia to search thousands of stars and hundreds of galaxies across multiple bands for technosignatures \cite{enriquez2017turbo, price2018blpdr, Price2020,price2021blpdr}. 
Despite the fact that  \textcolor{black}{these} radio telescopes are located in radio-quiet zones, Radio Frequency Interference (RFI) due to human technology still poses a major challenge for SETI research. In order to reject RFI,
one of the techniques employed by the BL team is that of spatial filtering using ``cadence'' observations, also known as ``position switching''.
 \textcolor{black}{The key idea is that an ETI signal observed on-axis in the primary beam should only appear in the ``ON-source'' scans, whereas RFI being near-field in nature would appear in multiple adjacent observations within a cadence irrespective of whether it is on or off source.}
However, the presence of RFI in observational data can often still result in a high false positive rate, as shown by previous searches on \textcolor{black}{a} similar GBT dataset as employed in this work, \textcolor{black}{where} 29 million hits \textcolor{black}{was} reported by \cite{enriquez2017turbo} and 37 million hits \textcolor{black}{was} reported by \cite{Price2020}.
 \textcolor{black}{In addition,} one can imagine a nearly infinite range of possible ETI signals which might not be captured by the  \textcolor{black}{conventional} de-Doppler  \textcolor{black}{SETI} algorithm \textsc{turboSETI} \cite{enriquez2017turbo,turboseti}.

Recently, Machine Learning (ML) has seen an increasing application in the field of astronomy\textcolor{black}{, thanks to its ability to generalize relationships in big datasets}. 
\textcolor{black}{In the context of SETI, some examples include} a generic signal classifier for observations obtained at the Allen Telescope Array \cite{Harp2019} and at the Five-hundred-meter Aperture Spherical Radio Telescope (FAST) \cite{Zhang_2020}, Convolutional Neural Networks (CNN) based RFI identifiers \cite{Pinchuk2021,CZECH201852}, as well as anomaly detection algorithms \cite{seti_self_super,Brzycki2020} \textcolor{black}{, although none of these work have yet to construct a purely ML-based SETI analysis pipeline.}
\(\beta\) is the hyperparameter that adjusts the weighting of the KL-Divergence Loss in a traditional VAE \cite{bvae}.
\textcolor{black}{Here we apply recent advances in disentangled deep learning, especially regarding the $\beta-$Variational Autoencoders ($\beta$-VAE) \cite{bvae} framework, in combination with a Random Forest decision tree to conduct the first comprehensive ML SETI.}
$\beta$-VAE defines a neural network that \textcolor{black}{implicitly learns} and identifies uncorrelated features within a dataset. 
This design ultimately tackles the black box problem with neural networks by forcing the network to learn  \textcolor{black}{human}-interpretable features from training data \cite{vae}. 
More concretely, this design of autoencoders allows the model to implicitly learn the features of ``cadence filtering'' and ``narrowband Doppler drifting signals'', meaning that a wider range of potential ETI signals can be searched. Furthermore, autoencoders can also implicitly learn how RFI appears in observational data, without having to be programmed for the nearly-infinite possible morphologies an RFI signal can take. 
\textcolor{black}{We apply our ML model} on the BL GBT 1.1--1.9\,GHz dataset of 820 nearby stars from 1004 cadences.
 We analyze the observational data in small snippets of 4096 frequency bins $\times$ 16 time bins,
which provides sensitivity to a maximum drift rate of $\approx\pm$10\,Hz/s (refer to the Methods section). 
We devise an overlapping search (Supplementary Fig.~\ref{fig:driftrange}) that \textcolor{black}{covers} snippets offset by half the size of a snippet window. 
Our search includes a total of 57 million unique snippets, excluding the regions of the band affected by instrumental effects.

\section*{The ML model and training architecture} \label{sec:ML}

\begin{figure}
\centering
\includegraphics[width=\textwidth]{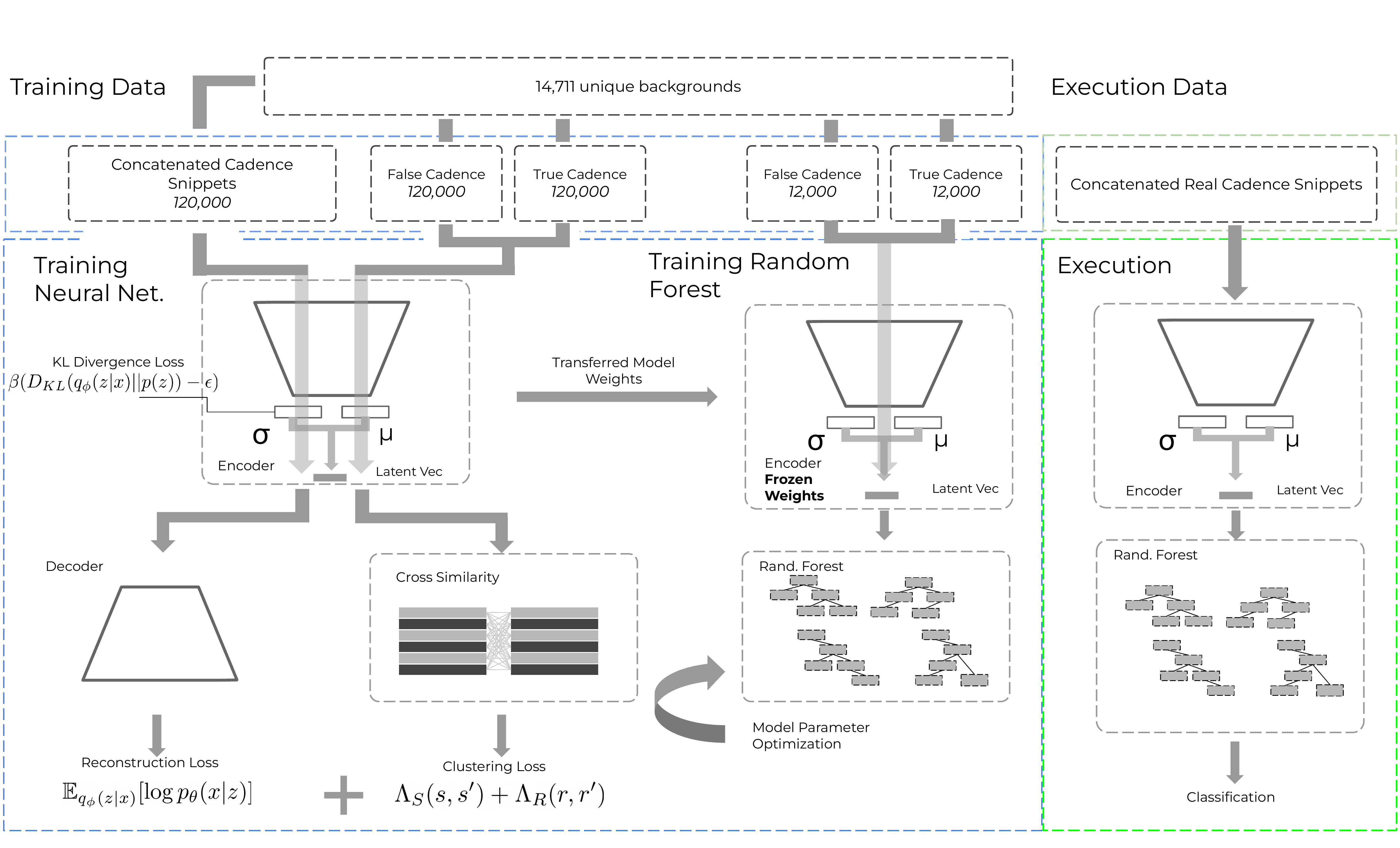}
\caption{Model training and execution scheme. Backward propagation of the neural network training is not shown for brevity. }
\label{fig:schema}
\end{figure}

Our encoder consists of \textcolor{black}{convolution layers with} 3$\times$3 kernels and a total of 8 layers with filter sizes of [16, 32, 64, 128] \cite{cnn}.
The number of convolutional layers \textcolor{black}{is} determined using a Bayesian optimization technique \cite{snoek2012practical}, whereas the filter sizes are fine-tuned empirically.  
We first stack these convolutional layers to extract ``spatial'' features from the spectrogram and subsequently feed the data through a traditional fully-connected neural network. Here the model splits off into two layers: the mean ($\mu$) and the log of the deviation ($\sigma$). These two are fed into a sampling layer as part of a standard encoder model \cite{vae}.
The decoder attempts to recreate the original spectrogram by performing the reverse of what the encoder did, as it is by definition inversely symmetrical to the encoder. The latent space in between the encoder and the decoder has a dense layer of 512 parameters and the latent vector is 8 dimensional in shape. The final model design is shown in Fig.~\ref{fig:schema}. It is implemented using  \textsc{Tensorflow}\footnote{https://www.tensorflow.org/} and \textsc{Keras}\footnote{https://keras.io/}.

The original VAE framework can be augmented with a single hyperparameter $\beta$ in the loss function that modulates the learning constraints applied to the model, specifically by controlling how much the model is penalized for how the model constructs the sample layer \cite{bvae}. For our model we empirically select $\beta$ = 1.5 \textcolor{black}{after comparing results from a number of validate sets}, which appears to best help the model to learn disentangled features in the latent space.
\textcolor{black}{
To further improve the model's performance, we introduce an inductive bias where we assume that ETI signals have similar features in the ON's and different features in the OFF's \cite{Inductivebias}. We also assume that RFI will have similar features in all the OFF's. This inductive bias is achieved by training the model to minimize or maximize an euclidean loss metric during training when presented with labeled data. Refer to the Methods section on Mathematical representations of the \(\beta\)-VAE ML Architecture for further details on the mathematical representations of our model.}

After constructing the $\beta-$VAE we create a Random Forest classifier \cite{randomforest} to perform the classification. Random Forests use an ensemble of decision trees to build a classifier, with each tree acting as a vote of confidence. We employ a Random Forest in combination with our $\beta$-VAE network because we needed to turn the encoder into a classifier and a Random Forest model is \textcolor{black}{fast to train and execute}. The greater the number of estimators (trees), the more robust a classifier is. We use the \textsc{SciPy} library\footnote{https://www.scipy.org} with 1000 estimators and bootstrap it when constructing the individual trees. Empirically this gives the best run-time vs performance metric. 

\begin{figure}[H]
\begin{subfigure}{.5\textwidth}
  \centering
  \includegraphics[width=\linewidth]{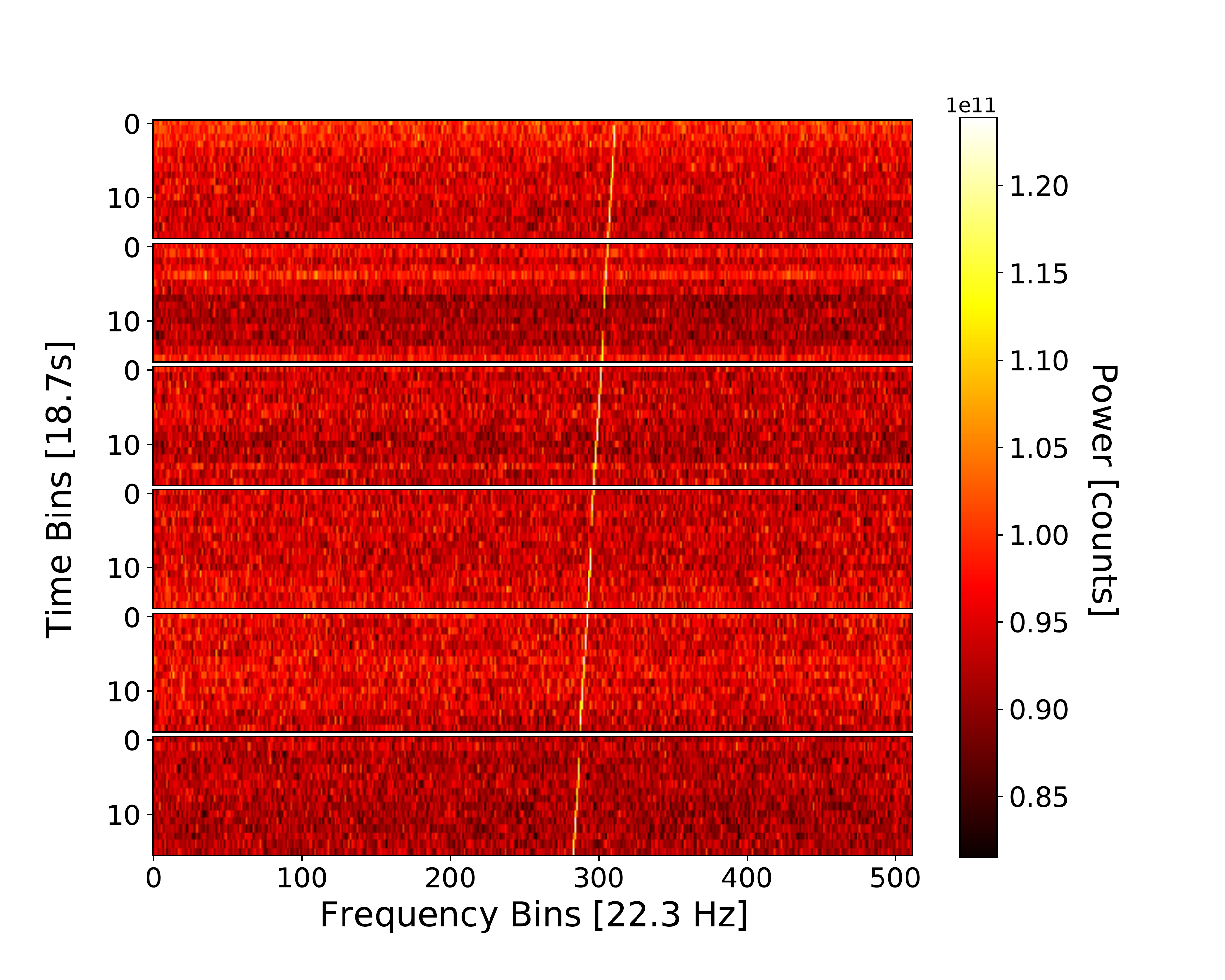}
  \caption{False with injected RFI}
  \label{fig:sfig1}
\end{subfigure}
\begin{subfigure}{.5\textwidth}
  \centering
  \includegraphics[width=\linewidth]{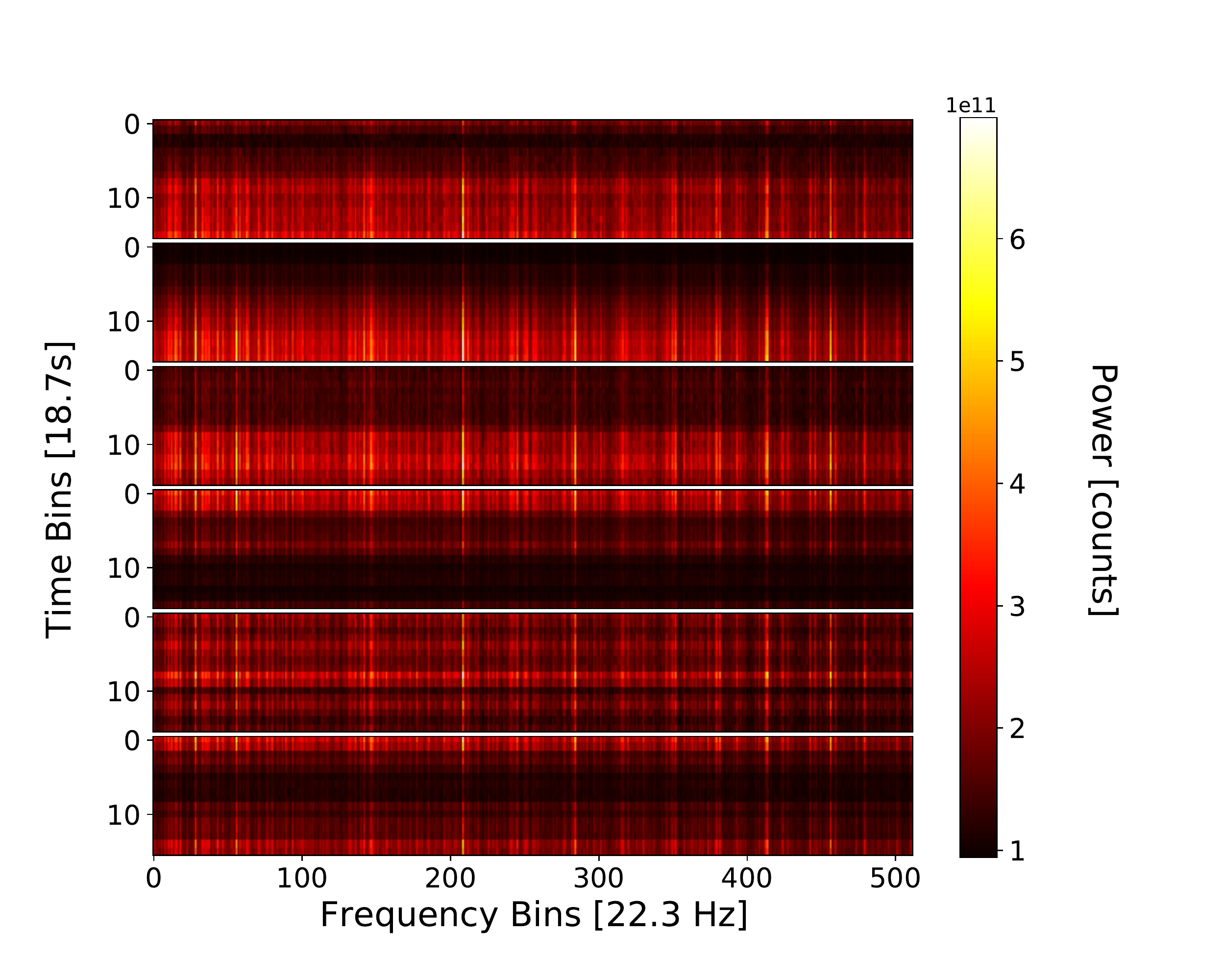}
  \caption{False with no injected RFI}
  \label{fig:sfig2}
\end{subfigure}
\begin{subfigure}{.5\textwidth}
  \centering
  \includegraphics[width=\linewidth]{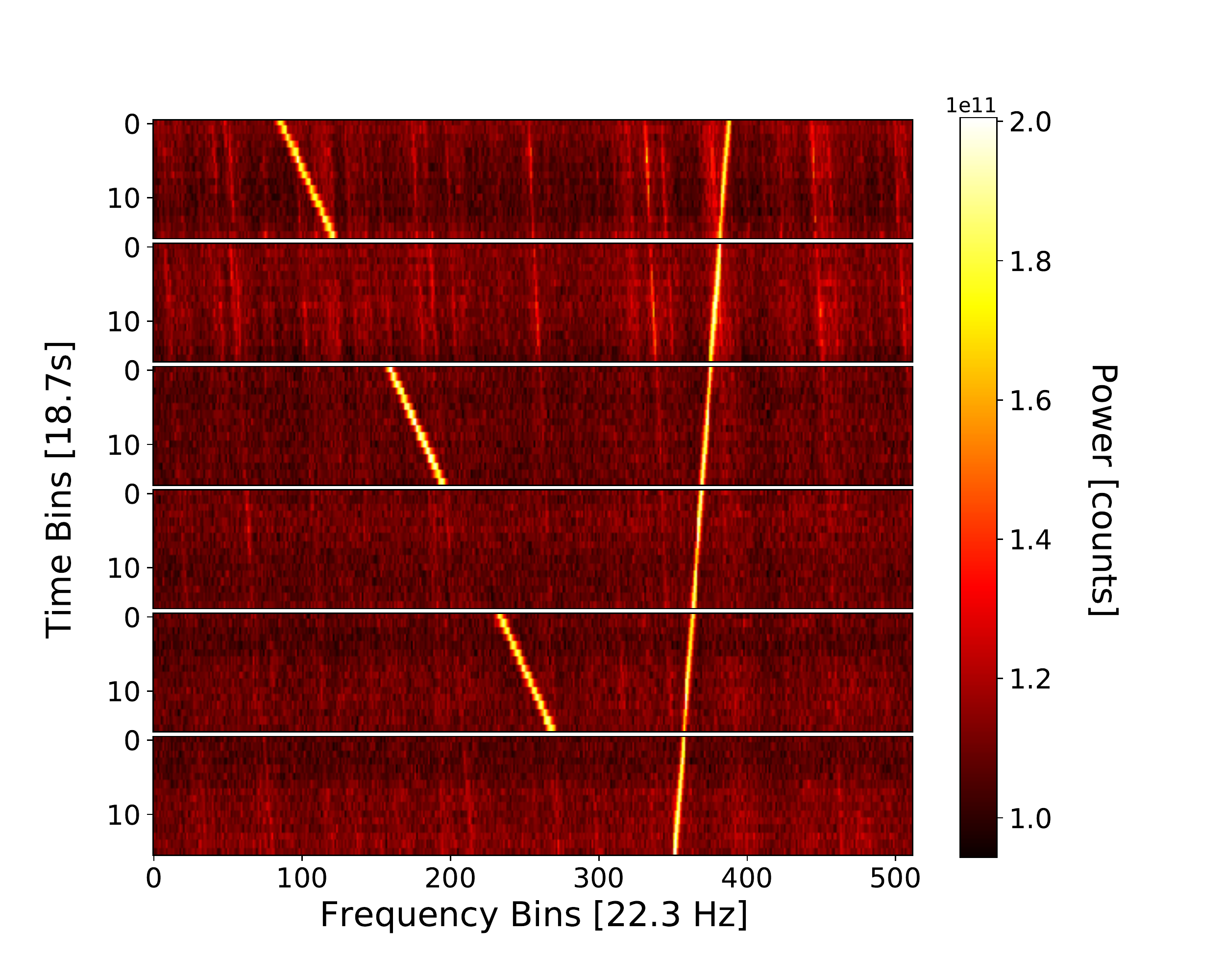}
  \caption{True with injected RFI}
  \label{fig:sfig3}
\end{subfigure}
\begin{subfigure}{.5\textwidth}
  \centering
  \includegraphics[width=\linewidth]{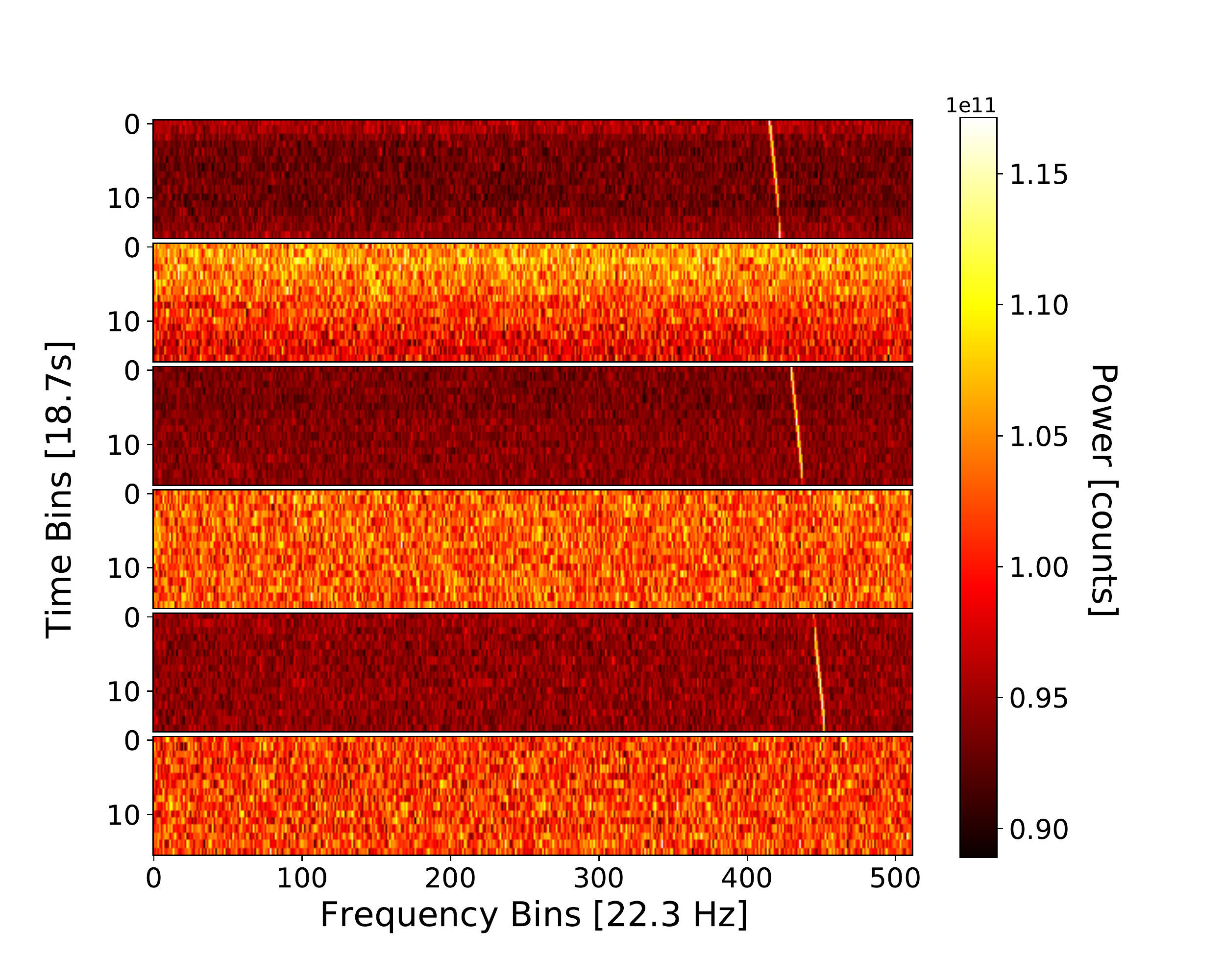}
  \caption{True with no injected RFI}
  \label{fig:sfig4}
\end{subfigure}
\caption{Examples showing the four types of training data. The bright streaks are signals detected. Each set of 6 panels represent 6 sequential ON-OFF observations. These spectrograms have a frequency resolution of 22.3\,Hz and a time resolution of 18.7\,s. \textcolor{black}{Both the injected Extraterrestrial Intelligent (ETI) Signal and Radio Frequency Interference (RFI) signals have an S/N of 20.}
}
\label{fig:4examples}
\end{figure}

In order to train our ML algorithm and evaluate the model, we need to provide labelled data for the model to learn from. There are three main categories of labelled data that are relevant: (1) False data with no ETI signals, (2) True data with ETI signals, and (3) True data with ETI signals and RFI. 
Since we do not have a sample of real observational ETI signals for this purpose, we generate simulated events by artificially injecting signals into the input spectrograms using the Python package \textsc{setigen}\footnote{https://github.com/bbrzycki/setigen}.
A total of $\NBACKGROUND$ different snippets of backgrounds are used, obtained from three different cadences (see the Supplementary Tables. 1 and 2 for more details). \textcolor{black}{This} large sample of backgrounds should provide a good variety of scenarios to help the generalization of the training model.
\textcolor{black}{From these backgrounds,} we randomly draw a heuristic 120,000 sample\textcolor{black}{s} to form the training set. For (1), we use one quarter (30,000 samples) of the original backgrounds without injections\textcolor{black}{,} and an additional 30,000 samples with RFI injected to the backgrounds. Another 30,000 samples are labelled as (2), where we inject ETI signals into the backgrounds and are referred to as ``True Single Shot'' in this paper. Finally, the last 30,000 samples are labelled as (3), where we inject both ETI signals and RFI with a 1:1 relative intensity ratio into the background spectrograms. 
We have not simulated all combinations of intensity ratios since the model will\textcolor{black}{,} in principle\textcolor{black}{,}  generalize this parameter. 
See Fig.~\ref{fig:4examples} for an example of these three types of labelled data used. 
A caveat is that if a true, non-synthetic ETI signal is present in the original backgrounds, it would lead to a mislabelling of our training dataset. This is an unlikely scenario, and our large, 120,000-sample training set should minimize the effect of individual mislabelled data. \textcolor{black}{Also note that we only train our model to detect signals that last for the entire cadence across the three ON scans. We do not attempt to assess the model's performance towards intermittent ETI signals or signals that have a shorter span than the full cadence.}

Another advantage of using simulated data is that we can control the injection parameters, namely the signal-to-noise ratio (S/N), the drift rate (DR) and the width ($w$) of the signal. We set a range of 10 to 50 for the S/N and a DR of $\pm$8\,Hz/s. Note that simulated ETI signals with zero drift rate are included. For $w$, we choose a range where \textcolor{black}{$\mathrm{DR}\times dt + 5 \leq w \leq \mathrm{DR}\times dt + 55$\,Hz, with $dt$ being the sampling rate.}
The lower bound offset of 5\,Hz is chosen to be slightly wider than the \textcolor{black}{amount of drift in one time bin at a given drift rate}.
This is necessary to prevent quantization of the injected signals as we do not have the native resolution in the filterbank-format observational data. 
Within the above ranges, we randomly draw a combination of S/N, drift rate and width for each simulated signal. See Supplementary Fig.~\ref{fig:simdata} for the distribution of the final set of injected parameters. 

In order to obtain the clustering loss matrix and the reconstruction loss vector, we need to train the encoder concurrently with concatenated (collapsed) cadences as well as non-collapsed cadences. We generate separate training sets for these to increase the diversity of the training set --- that is, we have 120,000 samples of concatenated cadences as well as an independent 120,000 samples of false cadences and another 120,000 samples of true cadences.
Refer to Fig.~\ref{fig:schema} for the breakdown of the size of the training set. 
The Random Forest model requires separate training since it needs to be built specific to a neural network where the weights are held constant.
This is achieved by taking the outputs encoded by the $\beta$-VAE and feeding the results into the Random Forest with synthetic labelled data. 
We create 12,000 false cadences and 12,000 true cadences (\textcolor{black}{24,000 total}) with  \textcolor{black}{the same} breakdown ratio of various labelled categories.
Overall, it takes roughly 12\,hr to generate and train the full model with 100 iterations each time on our machines (see the Supplementary Figure 6 for further details on the specs of these computers).

\section*{Model Evaluation and Demonstration} \label{sec:evaluation}
We test our algorithm against traditional ML approaches, including a support vector machine (SVM) \cite{svm}, a Random Forest \cite{randomforest}, a classic\textcolor{black}{, artificial} neural network \textcolor{black}{(ANN)}\cite{cnn}, and a 3-D convolutional neural network \textcolor{black}{(CNN)}\cite{cnn} \textcolor{black}{in order to demonstrate its performance. }
The SVM and the Random Forest are both classifiers and can only work with feature vectors as input, so we perform an additional step of applying principal component analysis (PCA) on the unrolled feature vectors.
We implement these models using the python module \textsc{Scipy}\footnote{https://www.scipy.org/}  with \textcolor{black}{further fine tuning discussed in sections "alternative model hyperparameter tuning" of the methods.}
\textcolor{black}{A test bench of labelled data using different backgrounds were generated, which} consists of \textcolor{black}{24,000} simulated snippets, with the same ratio of labelled events as our training set described \textcolor{black}{earlier}. The goal is to subject different ML models to this test bench and see how successful they are in recovering the correct signal classifications.  

\begin{figure}[H]
\centering
\includegraphics[scale=0.5]{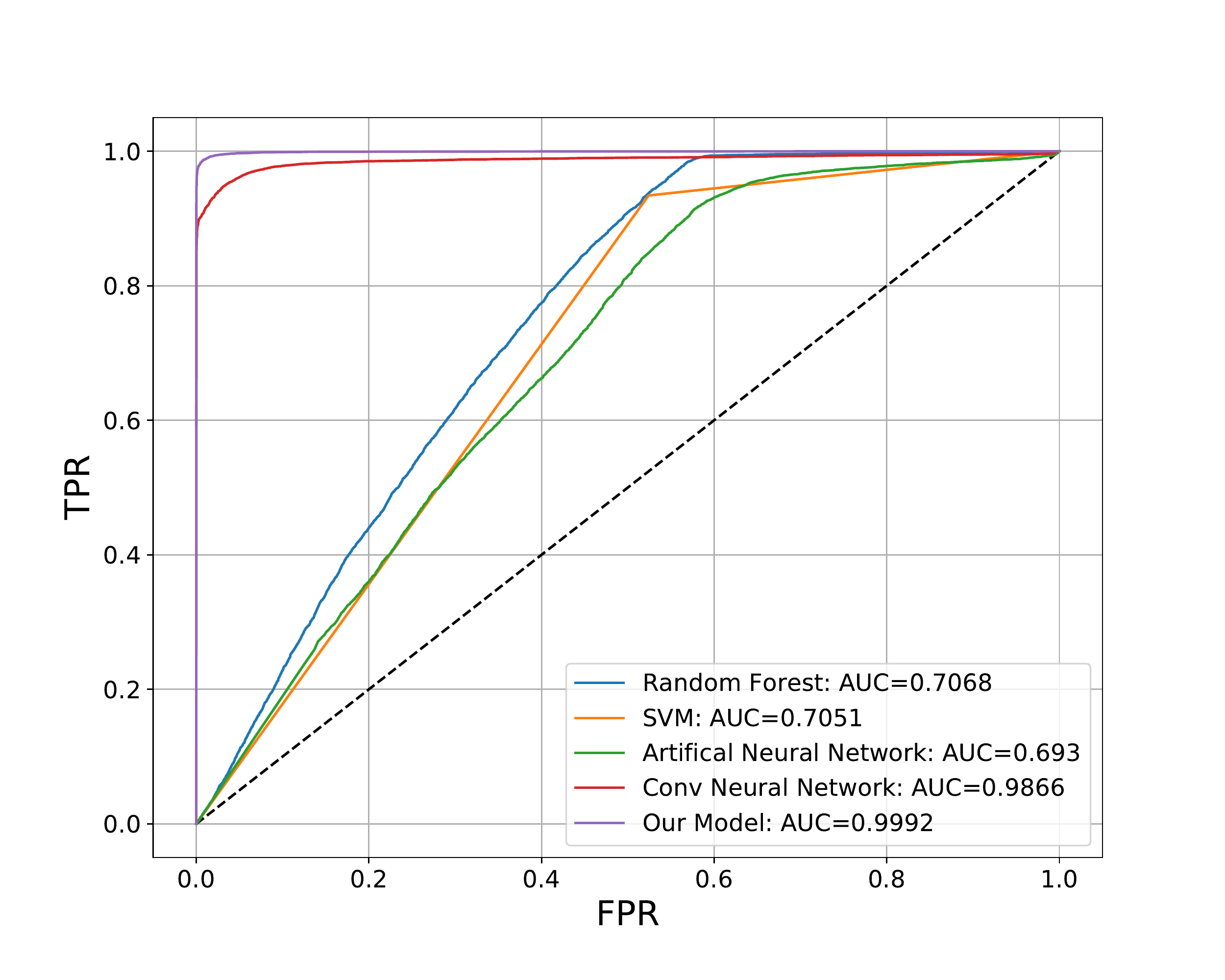}
\caption{An receiver operating characteristic curve (ROC) comparing the true positive rate (TPR) against the false positive rate (FPR) at various threshold settings for a number of ML models. The corresponding Area Under the Curve (AUCs) are listed in the legend. Models above the dashed line are considered better than random.} 
\label{fig:roc}
\end{figure}
Fig.~\ref{fig:roc} shows a receiver operating characteristic (ROC) curve which evaluates the performance of each model at varying classification thresholds.
All ML models perform well and are above the diagonal line. The Area Under the Curve (AUC) provides an aggregate measure of performance summing over all classification thresholds. A perfect model would have an AUC of 1. \textcolor{black}{Our ML model has} an AUC of \textcolor{black}{0.9993}, \textcolor{black}{a precision value of 0.993, a recall of 0.985 and an F1 score of 0.989. These are} significantly higher than  \textcolor{black}{most of} the other classical ML models\textcolor{black}{, and marginally better than the second best model of the CNN}. This provides evidence that simpler ML solutions perform less well than our $\beta$-VAE-based model.

\textcolor{black}{We also} evaluate the model performance across the \textcolor{black}{frequency band, the S/N and the drift rate ranges}. 
\textcolor{black}{From Supplementary Fig.~\ref{fig:evaluation-fullband}a, we find} that \textcolor{black}{the overall accuracy is} fairly high across the full band \textcolor{black}{at a probability threshold of 50$\%$}, with the majority of the sub-band having accuracy over 97.5\%. In other words, False, Single True and Double True labelled data were correctly identified  over 97.5\% of the time. The \textcolor{black}{few percent drop of accuracy} near 1600\,MHz corresponds to regions of the band that are heavily affected by RFI signals. As mentioned earlier, high false positive rate has been a persistent bottleneck in SETI efforts. Even a false positive rate of a few percent can translate to tens of thousands of candidates, reducing the search efficiency. We can attempt to drive down the false positive rate by selecting a higher classification threshold. 
\textcolor{black}{As shown in Supplementary Fig.~\ref{fig:evaluation-fullband}b, we find that at a threshold of 90\%, we have} a $<$1\% false positive rate  with a maximum 1-$\sigma$ uncertainty of $0.1\%$.
Naturally, a higher threshold will also mean worse true positive detection, meaning that we have a slightly higher chance of missing genuine ETI signals in the data. 
However, we see that the biggest drop in true positives happens in discarded regions of the band where prominent RFI is present. 
Barring those regions, we still have a true positive accuracy of above 95\% in the rest of the band.
For the S/N performance evaluation, 
we record a pretty consistent accuracy of $\gtrsim 95\%$ across S/N 20 to almost 80 \textcolor{black}{(Supplementary Fig.~\ref{fig:evaluation-SNR-DR}a)}, with the largest uncertainty of $\pm 2.7\%$ per S/N bin.
For the drift rate performance evaluation, 
\textcolor{black}{we find} that our performance drops as the drift rate increases \textcolor{black}{(Supplementary Fig.~\ref{fig:evaluation-SNR-DR}b)}. At 8\,Hz/s, we have roughly 70\% accuracy in identifying ``Single True'' injected signals. This drops to about 35\% at 10\,Hz/s. This degradation can be explained by our unbalanced test samples \textcolor{black}{(further discussed in the Methods section)}, where fewer high drift rate signals were employed in the training of the ML model.

\section*{Search Results} \label{sec:results}
\begin{figure}[H]
\includegraphics[scale=0.31]{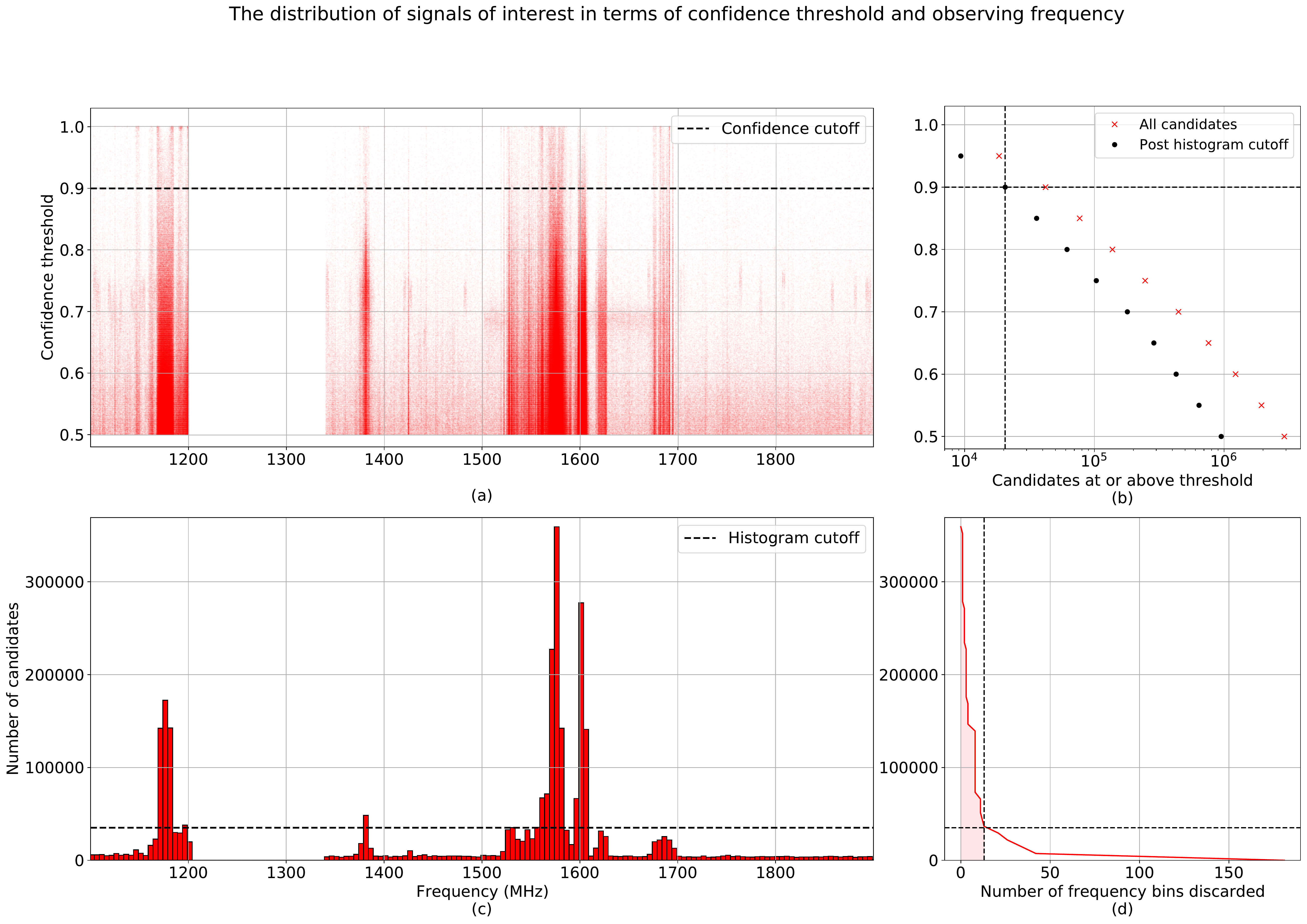}
\caption{(a) The distribution of \textcolor{black}{signals of interest} in terms of confidence threshold and observing frequency. (c) A histogram of the number of \textcolor{black}{signals of interest} at each observing frequency, with histogram bin size of 4.97\,MHz. (b) The number of \textcolor{black}{signals of interest} at or above a given classification threshold. The red data points are without the frequency histogram-based filtering, whereas the black data points are after such filtering. (d) The number of histogram bins being filtered out at progressively lower cutoff number of \textcolor{black}{signals of interest}. } \label{fig:Search_stats_accum}
\end{figure}

\textcolor{black}{From the 115 million snippet windows searched,} our ML model returns a total of 2,917,789 \textcolor{black}{signals of interest}. Small drift rate signal can potentially be seen by two adjacent overlapped snippets. Currently, each snippet is searched independently and we have not attempted to associate detections across them. \textcolor{black}{Hence this 3 million signals of interest is likely an over count.} \textcolor{black}{We note that nearly 10\% of these events came from a handful ($\sim$1\%) of cadences. These observations are heavily contaminated by RFI and skew the false positive rate of our analysis.}
We visualize the distribution of \textcolor{black}{all} events as a function of observing frequency as shown in Fig.~\ref{fig:Search_stats_accum}.
It can be seen that certain observing frequencies contain a much higher number of events compared to the others --- for example the region around 1600\,MHz. This overlaps with known RFI at the GBT site specifically from persistent GPS signals. Those regions of the spectrum are heavily contaminated by RFI and it would be challenging to detect anything apart from RFI in those frequencies. \textcolor{black}{Indeed, from our performance assessment we see that our ML model is, by a few percent, slightly less accurate in these RFI-contaminated frequencies. }
Using the frequency histogram in the bottom left panel of Fig.~\ref{fig:Search_stats_accum} which has a histogram bin size of 4.97\,MHz, we empirically determine a threshold to discard frequency bins with more than 35,000 events per bin, since this represents a conservative level where the main peaks of RFI clusters can be flagged. This equates to discarding 13 histogram bins, which is about 65\,MHz of the entire band.

\begin{figure}[H]
\includegraphics[scale=0.305]{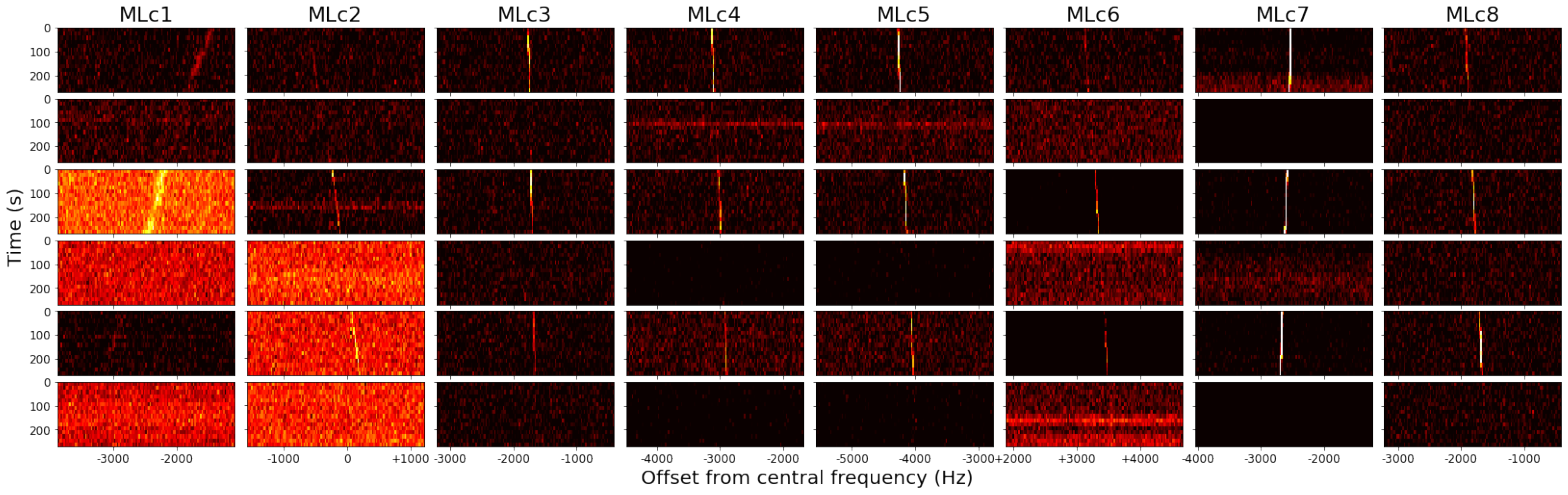}
\caption{\textcolor{black}{Waterfall plots of the $\NTOPCAND$ signals of interest. Each panel has a width of 2800\,Hz and the x-axis are referenced to the centre of the snippet where the signal is found, as reported in column~3 of Table~\ref{tab:top10}.}} \label{fig:8cands}
\end{figure}

The top right panel of Fig.~\ref{fig:Search_stats_accum} shows the number of \textcolor{black}{signals of interest} at a range of classification thresholds, which has a quadratically increasing trend towards low thresholds. We further limit our \textcolor{black}{signals of interest} to those that have a classification threshold of over 90\% in order to reduce the number of false positives. 
After applying these further filtering criteria, we are left with 20,515 \textcolor{black}{signals of interest} to assess.
Upon a visual inspection, we identify $\NTOPCAND$ promising \textcolor{black}{signals of interest} that show narrow, drifted signals in the three ON-scans (Fig.~\ref{fig:8cands}). Refer to Table~\ref{tab:top10} for their respective parameters. 
These $\NTOPCAND$ \textcolor{black}{signals of interest} come from five different stars. 
HIP~13402 and 54677 both have spectral type K whereas HIP~62207 is a G star, HIP~56802 an F star and HIP~118212 is an M star. They are all within 30 to 90\,ly from Earth.
All five targets were analyzed by \cite{enriquez2017turbo} and by \cite{Price2020} but they did not detect anything similar. 
\textcolor{black}{Re-observations of these five sources took place on May 21, 2022, at the Green Bank Telescope using the same set up as the original observations. We analyze these data using the same $\beta$-VAE pipeline and visually inspect all 72 events returned by the search. We do not detect anything like the  $\NTOPCAND$ signals of interest. This shows that no matter what the true nature of these signals are, they are not persistent in time. }
Given that the main goal of this work is to apply ML technique to identify signals with a specific pattern, we do not attempt to make a definite conclusion of whether these $\NTOPCAND$ signals are genuinely produced by ETI. 
\textcolor{black}{We encourage further re-observations of these targets.}

We individually compute the drift rate and S/N of these \textcolor{black}{signals of interest} as they are not a by-product of our ML pipeline. We estimate the drift rate in each ON-scan separately using \textsc{setigen}. The uncertainty of the drift rate is defined to be the deviation of drift rates among the three ON scans. 
A number of interesting events were rejected despite showing narrow band drifted signals as we find that the drift rates across the three ON-scans are not consistent. We note that a changing drift rate does not necessarily mean the variation is non-physical. 
\textcolor{black}{However, there is an infinite possibility of orbital motions that can phase connect three arbitrary drift rates. With a single dish SETI experiment, we have no independent way of verifying this and we consider it an intrinsic limitation of our search. We choose to exclude signals that are changing in drift rates for simplicity, as we cannot confidently relate them from one ON scan to the next. }
\textcolor{black}{We also point out that MLc4 and MLc5 have the same drift rates and come from an observation of the same target, although at slightly different times and at different parts of the band. }

\begin{table}[ht]
    \centering
        \resizebox{.8\textwidth}{!}{%
    \begin{tabular}{lcccccc}
\hline
ID & Target & Center Freq. & MJD & Confidence & Drift rate & S/N \\
 &  & (MHz) &  & &  (Hz/s) \\
\hline
MLc1	&	HIP\,13402	&	1188.539231	&	57541.68902	&	98.1	&	$+$1.11(25)	&  6.53 \\
MLc2	&	HIP\,118212	&	1347.862244	&	57752.78580	&	99.9	&	$-$0.44(7) 	&  16.38 \\
MLc3	&	HIP\,62207	&	1351.625410	&	57543.08647	&	93.7	&	$-$0.05(10) 	&  57.52\\
MLc4	&	HIP\,54677	&	1372.987594	&	57517.08789	&	99.9	&	$-$0.11(3) 	&  30.20\\
MLc5	&	HIP\,54677	&	1376.988694	&	57517.09628	&	97.9	&	$-$0.11(2)		&  44.58 \\
MLc6	&	HIP\,56802	&	1435.940307	&	57522.13197	&	99.9	&	$-$0.18(4) 	&  39.61 \\
MLc7	&	HIP\,13402	&	1487.482046	&	57544.51645	&	99.9	&	$+$0.10(2) 	&  129.16 \\
MLc8	&	HIP\,62207	&	1724.972561	&	57543.10165	&	99.9	&	$-$0.126(10) 	&  34.09 \\
\hline
\end{tabular}
}
\caption{The top $\NTOPCAND$ \textcolor{black}{signals of interest} identified by our ML model. We list the name of the ON-source target star, the frequency at the middle of the relevant $\sim$11\,kHz snippet, the MJD of the detection, the topocentric drift rate, as well as the S/N of the signal. The HIP prefix in the source name shows that these targets are drawn from the Hipparcos catalog \cite{HIP}.}
\label{tab:top10}
\end{table}

\textcolor{black}{Regarding the nature of the rest of the events, most of them look like false positives associated with RFI signals. We have not exhaustively classified every single event and it is entirely possible that there are additional ETI-like signals that we have not picked up, including those that are too weak to be seen by human eyes. 
Other reasons as to why the false positive rate is higher than in our validation process include overfitting of backgrounds. The diversity of the RFI environment and the brightness of the RFI signals mean that it is unavoidable that our model would sometimes focus more on the background than the foreground signals. In addition, we do not have real ETI signals to train the model on, and instead have to rely on simulated signals that are injected to the sky backgrounds. Injection of signals might have altered the statistics of the snippets, introducing artefacts that were unintentionally learnt by the ML model.}

\section*{Discussion} \label{sec:discussion}
This work represents the most comprehensive ML-based technosignature search to date, and improves on previous work by finding signals of interest not detected before\cite{enriquez2017turbo,Price2020}. We generate synthetic, labelled data to train a $\beta$-VAE framework together with a Random Forest classifier. We observe some level of generalization in the trained model, as the latent space shows interpretable features and the model is able to correctly identify signals beyond the initial training parameter space.

Overall we see a high degree of accuracy when the model is subjected to a different test bench, both across the frequency band as well as over a wide range of signal S/N. Our model also performs better in comparison to a number of classical ML models tested. 

By analyzing $\NCADENCES$ cadences corresponding to approximately 115 million snippets recorded with the GBT 1.1--1.9\,GHz receiver, our model returns 2.9 million events. Upon further filtering by discarding RFI affected frequency bins as well as by limiting the classification threshold to 90\%, we further reduce the number of events to approximately 20,000. By visually inspecting the individual diagnostic plots, we discover $\NTOPCAND$ promising SETI \textcolor{black}{signals of interest} with narrow band, drifted signals showing the expected on-off pattern \textcolor{black}{that were not identified by previous \textsc{turboSETI} searches} \cite{enriquez2017turbo,Price2020}. 
\textcolor{black}{We limit our search to drifting signals that have a uniform drift rate and are persistent across the three ON scans. However, a modified ML model can be adapted and re-trained to look for other ETI morphologies in the future, such as those with a spread spectrum emission. }
\textcolor{black}{One main} limitation of our ML implementation is that we have a non-uniform training set with fewer synthetic signals at higher drift rates (see Supplementary Fig.~\ref{fig:simdata}). 
This has likely skewed our model and \textcolor{black}{misguided} it that low drift rate events are more common.

Looking ahead, we hope to expand this ML technique to other Breakthrough Listen datasets to further increase the impact of ML on SETI. This includes other GBT and Parkes data, as well as the upcoming MeerKAT \cite{MeerKAT}, Very Large Array (VLA)\cite{COSMIC}, \textcolor{black}{Square Kilometre Array\cite{Siemion2015} and the next generation VLA (ngVLA; Ng et al., submitted)} interferometric SETI projects.\cite{Pinchuk_2022}

\bibliographystyle{naturemag}
\bibliography{references.bib}
\newpage

\section*{Methods}
\subsection*{The GBT 1.1--1.9\,GHz dataset} \label{sec:data}

We have chosen to deploy an ML SETI algorithm on the Breakthrough Listen GBT 1.1--1.9\,GHz dataset, sourced from several different observational campaigns, as it is one of the largest homogeneous SETI datasets available in a single location at the Berkeley SETI Research Center. The majority of this dataset has been made publicly available\footnote{https://seti.berkeley.edu/opendata}.
The dataset used in this work has a frequency range of  1023$-$1926\,MHz and consists of a total of $\NCADENCES$ cadences from $\NSTARS$ unique targets observed over 480\,hr. In total, $\sim 120$\,TB of data has been analyzed in this work.

Each cadence has six 4.8-min observations recorded in HDF5 format.  
In the analysis described here, we exclusively work with the fine-frequency resolution data\cite{lebofsky2019}. More than half of the cadences have a frequency resolution of 2.79\,Hz and 322,961,408 frequency channels. The remaining 397 
 cadences taken before 2016 April have a frequency resolution of 2.84\,Hz and 318,230,528 channels per observation.
All data have a time resolution of $\sim$18\,s and 16 time bins for each observation within the cadence. 
Parts of the band are affected by instrumental artifacts, including steep roll-off on the band edges and at the boundaries of an analog notch filter. These regions are at frequencies $\nu <$ 1.1\,GHz, $\nu >$ 1.9\,GHz and 1.2 $< \nu <$ 1.34\,GHz. Together these equate to about 30\% of the full band and are excluded from our search.

Two types of cadences have been used in the GBT observations, the ABACAD pattern and the ABABAB pattern. The difference between these is that the ``ON-source'' scans (the A scans) can be interspersed with identical ``OFF-source'' targets (the B scans) or three different ``OFF-source'' targets (B, C, D). In both cases, the minimum angular offset between ON and OFF source scans is at least six beamwidths, which is well beyond the primary and side lobes of the GBT beams. In fact, the majority (over 80\%) of the cadences analyzed have an angular offset of as much as 38 beamwidths.
We treat both types of cadences the same way in our analysis, since for the purpose of spatial filtering we only care that the ``OFF-scans'' are sufficiently far from the ``ON-scans''. 

\textcolor{black}{In our simulated data, we have assumed no time delay between the six observations in a cadence, although in reality, there is a small amount of slew time switching between ON and OFF scans. } We have discarded cadences that have a long ($\ge$2\,min) pause between the individual observations,
since that would significantly alter the data characteristics due to longer term variation in telescope performance and our ML model has not been trained to detect those.
\textcolor{black}{The $\NCADENCES$ analyzed have typically a small time gap} of the order of a minute \textcolor{black}{and we have effectively assumed the observations are continuous for simplicity.} 
For signals with a low drift rate of say
1\,Hz/s, this is not really an issue, since a minute of time delay would correspond to a shift of about 60\,Hz which is only 2 or 3 frequency bins in the downsampled snippet used for the ML search. However, this issue becomes more prominent at high drift
rates. For example, at the maximum detectable drift rate of 10\,Hz/s, during the minute of gap the signal would have shifted by
about 600\,Hz, which is roughly 25 downsampled frequency bins.
\textcolor{black}{We note that it is an oversight on our part, as not taking time delay into account could have affected our ability in detecting signals with high drift rates. }

\subsection*{Data preprocessing} \label{sec:training}
\begin{figure}[H]
\begin{subfigure}{0.75\textwidth}
  \centering
  \includegraphics[width=1.10\linewidth]{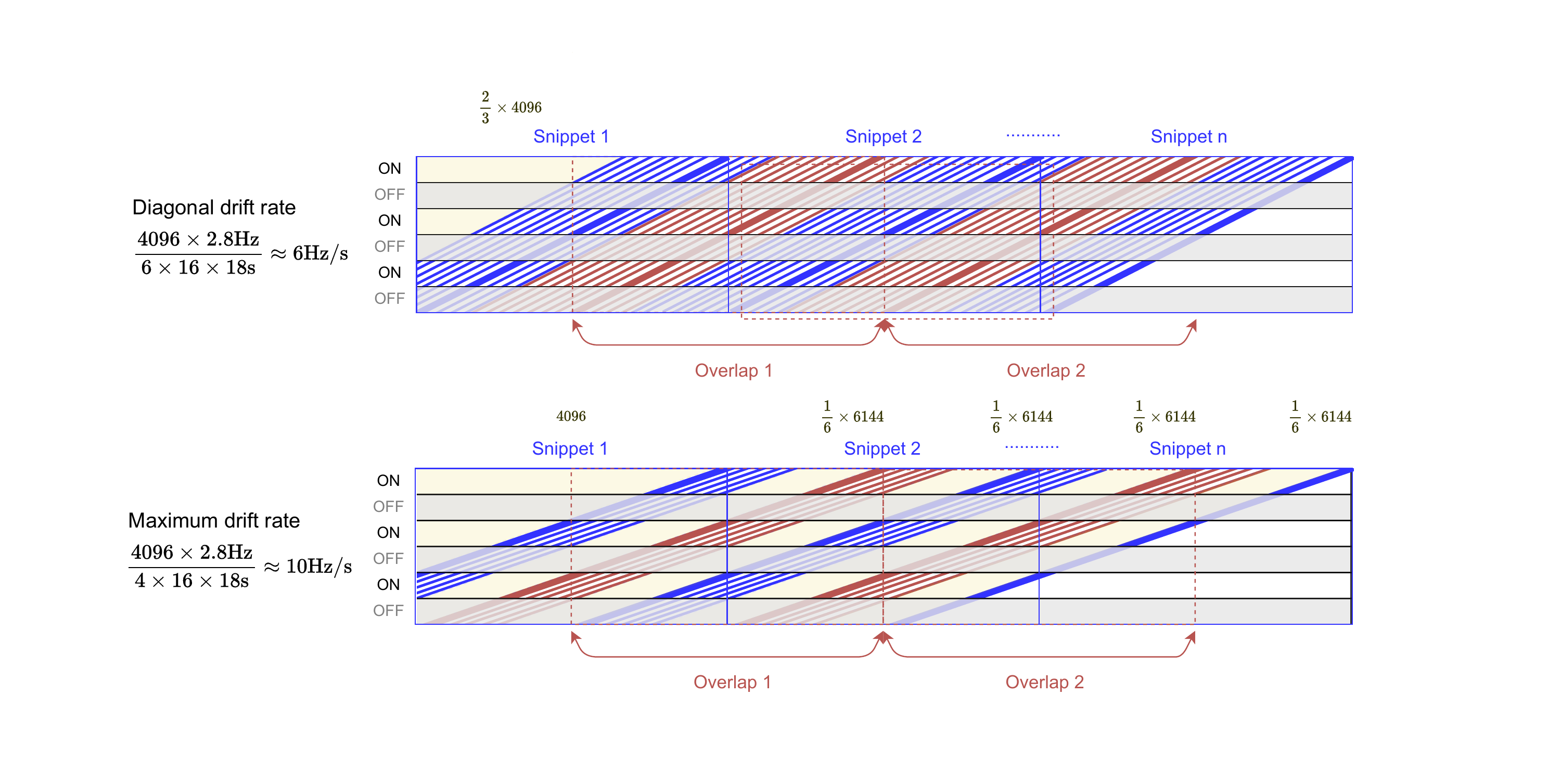}

\end{subfigure}%
\begin{subfigure}{0.25\textwidth}
  \centering
  \includegraphics[width=0.98\linewidth]{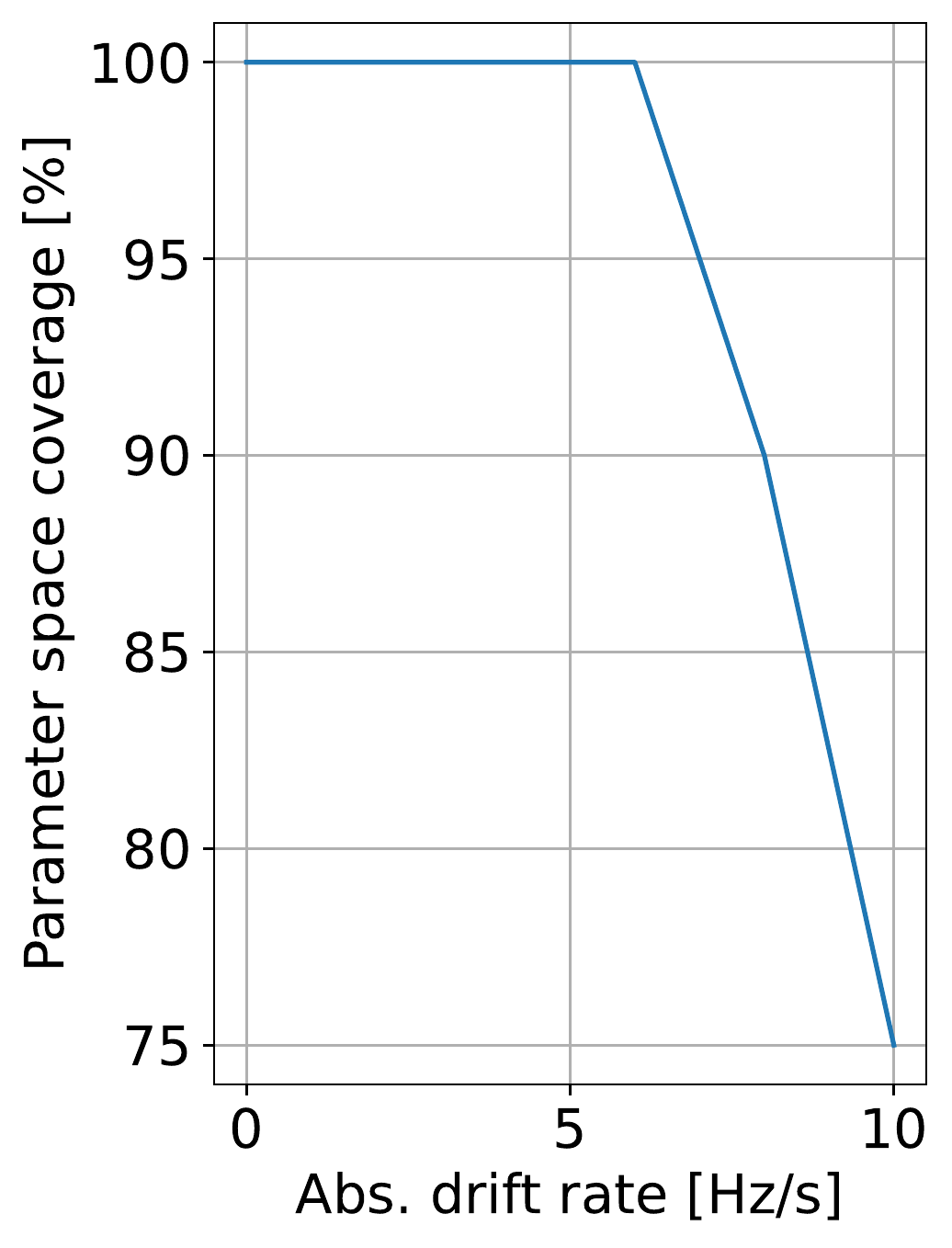}
\end{subfigure}
\caption{(a) A sketch of the overlap search method. Only positive drifting signals are shown for brevity as the negative drifts are symmetric. The top panel shows the diagonal drift rate at 6\,Hz/s. The blue drifting signals are detectable by the regular snippets (blue), whereas the red drifting signals in between are detectable by the overlap search (red). A small triangular region on the far left is not detectable by our search because any drifting signal starting there will have part of its pattern before the start of the first snippet, in the sense that not all three ``ON'' scans contain a signal. This triangular region exists for all non-zero drift rates and gets progressively larger. But in all cases, this represents a tiny region out of the many snippets within a cadence and thus we have practically \textcolor{black}{full} coverage for any drift rate below the diagonal drift.
The bottom panel shows the scenario with the maximum detectable drift rate at 10\,Hz/s. For example, this is when the first blue drifted signal barely shows up in the third ``ON'' scan. In this case, apart from the triangular region on the left, there will be additional gaps throughout the cadence that are not caught by neither the snippets nor the overlap search. The number of frequency channels missed are indicated on top of each of the yellow-highlighted regions.
(b) The corresponding parameter space coverage as a function of the absolute drift rate. We have practically full detectability up to the diagonal drift rate of 6\,Hz/s. Beyond that, the coverage drops to 75\% at the maximum drift rate of 10\,Hz/s.} 
\label{fig:driftrange}
\end{figure}

Neural networks perform better if the input data has a shape close to that of a CNN kernel which typically has a 1:1 aspect ratio. In contrast, the fine-resolution GBT data have a very elongated dimension, with over 300 million frequency bins by 16 time bins per observation. 
Indexing small partitions of the frequency band and searching them independently can solve that problem, but a small frequency band will reduce the maximum drift rates our algorithm is sensitive to, since the algorithm cannot detect any signal that drifts beyond the diagonal of the input data shape. Here we define the maximum drift rate detectable by a search to be when a signal starts drifting from the corner of the first observation and drift out near the opposite corner of the third ON scan, spanning a total of four observations (ON-OFF-ON-OFF) worth of time (see bottom panel of \textcolor{black}{Supplementary} Fig.~\ref{fig:driftrange}a). 

We find a good compromise in using snippets of 4096 frequency bins, which provides sensitivity to a maximum drift rate of 4096$\times$2.8\,Hz over 16$\times$18\,s per observation over 4\,observations $\approx\pm$10\,Hz/s. 
These higher drift rates have a proportionally higher chance to drift out of at least one of the snippets depending on where the signals start, partially altering the expected ON-OFF pattern and reducing the sensitivity of our model towards these cases. 
In order to alleviate this potential issue, we devise an overlapping search that will cover snippets offset by 2048 frequency bins which is half the size of a snippet window (\textcolor{black}{Supplementary} Fig.~\ref{fig:driftrange}). 
In the scheme illustrated in the top panel of \textcolor{black}{Supplementary} Fig.~\ref{fig:driftrange}a, any drift rate lower than the diagonal drift rate, i.e. 4096$\times$2.8\,Hz/(6$\times$16$\times$18\,s)$\approx$6\,Hz/s will either be detectable by the blue snippets or by the red overlap search. Only a very small yellow-highlighted region on the far left is missed since too much of the drifting signals happen before the start of the snippet. Compared to the large number of snippets per cadence, this equates to a 99.999\% of parameter space coverage.
For signals with higher drift rates than the diagonal, the overlap search would not be quite enough to fully cover some of the parameter space. The bottom panel of \textcolor{black}{Supplementary} Fig.~\ref{fig:driftrange}a shows the most extreme case at the maximum detectable drift rate of 10\,Hz/s. At this point, apart from a slightly larger triangle on the left that will be missed, there are now gaps between the blue and the red drifting signals that we are not sensitive to.
The probability of having the signal detectable within a cadence drops to $\sim$75\,\% at $\pm$10\,Hz/s (\textcolor{black}{Supplementary} Fig.~\ref{fig:driftrange}b).

We further downsample the snippets in frequency by a factor of 8 (to a frequency resolution of 22.32\,Hz or 22.72\,Hz) \textcolor{black}{by averaging every 8 pixels. This} results in an input spectrogram with shape  [time, frequency]=[16, 512]. This helps to further \textcolor{black}{improve} the training time and performance of the neural network, at the expense of decreasing the S/N of a potential signal contained in one channel by $\sqrt8$.

The dynamic range of these GBT data can vary by as much as a factor of $10^{11}$\textcolor{black}{, as a result of the long Fast Fourier Transform (FFT) used to generate these high spectral resolution data}. After the injection stage, we log normalize the data,  shift everything to be positive and we scale the data to have a final range within 0 and 1. Note that the normalization is done per snippet of each cadence and not across the entire cadence.

\subsection*{Drift rate range trained}
\begin{figure}
  \centering
  \includegraphics[width=0.9\linewidth]{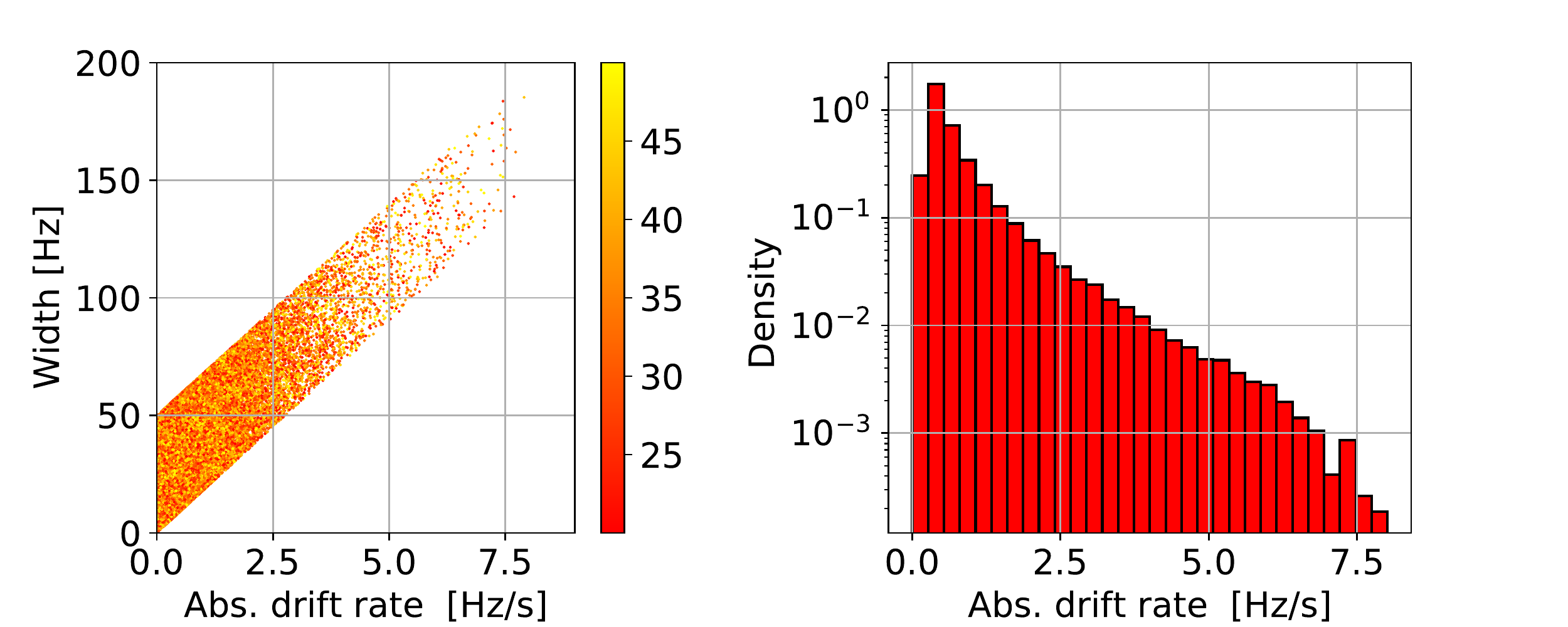}
\caption{(left) The distribution of widths vs drift rates for a randomly picked sub-set of 5000 synthetic signals generated for the training set. The colors represent the S/N of each signal.
There are fewer samples with higher drift rates as we discard everything that drifts partially out of frame.
(right) The corresponding density distribution of our simulated sample as a function of absolute drift rate.} 
\label{fig:simdata}
\end{figure}

Within the \textcolor{black}{S/N, width, and drift range} ranges \textcolor{black}{specified in Table~\ref{tab:background-training}}, we randomly draw a combination of S/N, drift rate and width for each simulated signal. 
See \textcolor{black}{Supplementary} Fig.~\ref{fig:simdata} for the distribution of the final set of injected parameters. 
Note that there are fewer simulated signals at higher drift rates. This is partly because there is a higher chance for them to drift out of our detectable parameter space as can be seen in \textcolor{black}{Supplementary} Fig.~\ref{fig:driftrange}, and also because we have not simulated signals that start outside of a snippet but drift into it. \textcolor{black}{Supplementary} Fig.~\ref{fig:simdata}b shows the density distribution of our simulated sample as a function of drift rate. 
To quantify the overall detectability of this ML model as a function of drift rate, we should take into account both the parameter coverage as provided by the overlap search (right panel of \textcolor{black}{Supplementary} Fig.~\ref{fig:driftrange}), as well as the sampling distribution of simulated data (right panel of \textcolor{black}{Supplementary} Fig.~\ref{fig:simdata}).
We estimate that we have a $\sim$80\% sensitivity towards a 6\,Hz/s signal that was present in this dataset, a $\sim$60\% sensitivity towards \textcolor{black}{an} 8\,Hz/s signal and a $\sim$25\% sensitivity towards a 10\,Hz/s signal.
The fact that we do not have a balanced test class with respect to drift rate means our model is less well-trained for the high-drift region of the parameter space.

\subsection*{Mathematical representations of the $\beta$-VAE ML Architecture}
%Overview of VAE and autoencoder
In this analysis\textcolor{black}{,} we use a VAE-based ML model \cite{vae}, which is one of two main families of deep generative models that can produce highly realistic versions of the input data. Autoencoders are neural networks architectures with an encoder and a decoder that form a bottleneck for the data to go through. They are trained to minimize the loss of information during the encoding-decoding process, which is achieved by iterations of gradient descent with the goal of diminishing the reconstruction error. A VAE is a special kind of autoencoder whose encoding distribution is regularized. The term ``variational'' reflects the relationship between the regularization and the variational inference method in statistics.
%Introducing the beta parameter
 A mathematical representation of our model is shown by Equation~\ref{eq:betaVAE}, which is a slightly modified version of Equation~3 in \cite{vae}. In the referenced paper, \(\mathcal{L}\) was used to represent the Lagrangian. Here we flip the sign of the equation to show the loss function $L$ as follow:
\begin{equation}
L(\theta,\phi,\beta; x,z) = \overbrace{-\mathbb{E}_{q_\phi(z|x)}[\log p_\theta(x|z)]}^{\mathrm{Reconstruction~Loss}} +
\overbrace{\beta (D_\mathrm{KL}( q_\phi(z|x)||p(z)))}^{\mathrm{KL~Divergence}} .
\centering
\label{eq:betaVAE}
\end{equation}
In this Equation, $x$ is the observational data and we want to develop an unsupervised deep generative model which will learn the joint distribution of the data $x$ and a set of generative latent factors $z$.
This effectively is the probabilistic decoder, $p_\theta(x|z)$, with generative model parameters $\theta$, where for a given $z$ it produces a distribution over the possible corresponding values of $x$. Inversely, $q_\phi(z|x)$ is the probabilistic encoder with recognition model parameters $\phi$, where for a given data point $x$ it produces a distribution over the possible values of $z$ from which the data point $x$ could have been generated.
$\mathbb{E}_{q_\phi(z|x)}$ computes the (log-)likelihood of effectively each pixel that the model reconstructs. The function \(D_{KL}\) is the Kulback-Leibler (KL) Divergence.
To summarize, the two main terms in this Equation are the reconstruction loss and the $\beta$-weighted Kulback-Leibler (KL) divergence. The KL divergence metric is used to measure the statistical similarities between the mean and the log-normalized standard deviation layers. 

Specifically for our analysis, we also want the model to be able to group observations with similar features together in the model's latent space.  In order to achieve this we incorporate a clustering loss which we denote $\Lambda$. 
We use  $\Lambda_\mathrm{S}$ for SETI and $\Lambda_\mathrm{R}$ for RFI signals. 
We compute the latent vectors of the cadence for true SETI cases called $S$ with corresponding latent factor $s'$ and an input of $s$. Similarly, we label the corresponding parameters for false SETI cases (RFI) as $R$ with corresponding latent factor $r'$ and an input of $r$: 
\begin{equation}
S = q_{\phi}(s'|s) \,, \quad R = q_{\phi}(r'|r) \,.
\end{equation}
Each parameter represents a set of six feature vectors corresponding to the six observations each encoded from the cadence.
We define two lists to index each vector in the set of features as follow:
\begin{equation}
\text{ON}= \{0,2,4\} \,, \quad 
\text{OFF}= \{1,3,5\} \,.
\end{equation}
We compute the distance between each encoded feature vector on the true SETI cadence we have for every pair of ON targets (3$\times$2 different combinations) and pair of OFF targets (another 3$\times$2 different combinations). We compute their similarities, whereas for a pair of adjacent ON-OFF (3$\times$3 different combinations) we compute the inverse. The indices $i$ and $j$ represent the ID of the observation and each goes from 0 to 5. The only condition is that $i \neq j$.
We then average over $n = 21$ combinations to get the clustering losses:
\begin{equation}
\Lambda_S=\frac{1}{n}(\sum_{i\in \text{ON}}\sum_{ \substack{j \in \text{ON}}} ||S_i-S_j||^2 + 
\sum_{i\in \text{OFF}}\sum_{ \substack{j \in \text{OFF}}}||S_i-S_j||^2  + \sum_{i\in \text{ON}}\sum_{j \in \text{OFF}} 
\frac{1}{||S_i-S_j||^2}) \,.
\end{equation}
For the false SETI (RFI) cadence, since we inject a continuous signal across all six observations, we can assume that every pair of observations will look similar. Therefore, we compute the distance between each encoded feature vector for the RFI cadence by calculating the similarity between every pair and average over them to get the clustering losses:
\begin{equation}
\Lambda_R=\frac{1}{n}(\sum_{i\in \text{ON}}\sum_{ \substack{j \in \text{ON}}} ||R_i-R_j||^2 + 
\sum_{i\in \text{OFF}}\sum_{ \substack{j \in \text{OFF}\\ }}||R_i-R_j||^2  + \sum_{i\in \text{ON}}\sum_{j \in \text{OFF}} 
||R_i-R_j||^2 ) \,.
\end{equation}

\textcolor{black}{ A smaller \(\Lambda_S\) implies that all the ON targets have vectors that are close together and all the OFF targets are close together, whereas the ONs and OFFs are far apart. This is in contrast to a smaller \(\Lambda_R\) where all the signals are close to each other.} 
In other words, we can identify true ETI signals this way because the ON scans (A) will be separated furthest from the rest of the OFF scans (B, C, D). 

By formulating the cost function in this manner, we help the neural network to make an ``implicit'' classification that groups relevant features together. Finally we include another hyperparameter weight $\alpha$ to tune the relative ``importance'' of the loss metric. We empirically set $\alpha = 10$ as it gave the best performing model. In summary, the loss function ($L$) of our model can be represented by: 
\begin{equation}
L(\theta,\phi,\beta; x,z,s,r) = \overbrace{-\mathbb{E}_{q_\phi(z|x)}[\log p_\theta(x|z)]}^{\text{Reconstruction Loss}}
+ \overbrace{\beta (D_{KL}(q_\phi(z|x)||p(z)))}^{\text{KL Divergence }} + \alpha  
\overbrace{(\Lambda_S(s,s')}^{\text{SETI }} + \overbrace{\Lambda_{R}(r, r')}^{\text{RFI  }}) \,.
\end{equation}

\subsection*{Backgrounds used for the Training Data and Test Benches}
 For the training data, a total of $\NBACKGROUND$ different snippets of backgrounds are used, obtained from three different cadences (see \textcolor{black}{Supplementary} Table~\ref{tab:background-training}). These backgrounds are taken from regions of the band (1418.7 to 1587.9\,MHz) that overall showed higher than average RFI as determined by the previous \textsc{turboSETI} search \cite{enriquez2017turbo}, which we can confirm by visually inspecting these regions in the three cadences. The intention is that our model is trained with some examples of the most RFI-contaminated observations. A more rigorous selection of backgrounds could be achieved by identifying snippets with statistical anomalies in energy, as well as by employing a larger number of cadences to increase the diversity in the backgrounds. However, we believe that our large sample of backgrounds should provide a good variety of scenarios to help the generalization of the training model.

\begin{table}[ht]
    \centering
        \resizebox{\textwidth}{!}{%
    \begin{tabular}{ccccccc}
\hline
Training target & Cadence & Freq range & Bandwidth & Freq. res & No. of snippets & No. of test samples\\
  & & (MHz) &  (MHz) &  (Hz) \\
\hline
\multirow{3}{*}{$\beta$-VAE} & HIP\,110750 & 1418.7$-$1475.1 & 56.4 & 2.79 & 4928 & \multirow{3}{*}{120,000}\\
 & HIP\,13402  & 1475.1$-$1531.5 & 56.4 & 2.83 & 4855 \\
 & HIP\,8497  & 1531.5$-$1587.9 & 56.4  & 2.79 & 4928 \\
 Random Forest & HIP\,110750 & 1418.7$-$1475.1 & 57.139 & 2.79 & 4928 &  24,000 \\
\hline
\end{tabular}
}
\caption{The breakdown of \textcolor{black}{the} number of background snippets and the frequency ranges of each cadence used for the training dataset. The number of test samples simulated in each training set is also listed.}
\label{tab:background-training}
\end{table}

In order to evaluate the model performance, we generate \textcolor{black}{different} test benches of labelled data.  \textcolor{black}{Supplementary} Table~\ref{tab:background-testbench} summarizes the properties of these test sets.
For the test where we compare our \(\beta\) -VAE model with other ML models, we create a test bench that consists of an RFI-free band (1418.7$-$1475.1\,MHz) of a new observation of HIP15638 that our model has not seen before, as well as an RFI\-affected band (1644.5$−$1700.7\,MHz) from HIP\,13402. 24,000 samples are drawn, with the same ratio of labelled events as our training set, i.e. half False data, a quarter ``True Single Shot'' and a quarter ``True Double Shot''. The goal is to subject different ML models to this test bench and see how successful they are in recovering the correct signal classifications. 

In order to evaluate the model performance across the frequency band and across the S/N range, we create one test bench using an observation of HIP\,15638 as the background.
We split the full GBT data into 16 sub-bands and we step through the S/N range of 20 to 80 with a step size of 2, resulting in 30 S/N bins. We deliberately simulate signals brighter than the max S/N of 50 used in the initial training dataset. The idea is to see if our model has successfully generalized to high S/N signals that it was not previously trained on. 
For each frequency sub-band and each S/N bin, we simulate 1000 test samples. This amounts to a total of 16$\times$30$\times$1000=\textcolor{black}{48,0000} simulated signals, again with the same ratio of labelled events and same ranges of widths and drift rates as our training set. 

For the frequency test, we average over all S/N and compare across the 16 sub-bands. 
Since our test bench consists of balanced data, i.e., \textcolor{black}{an} equal amount of False and True signals, we \textcolor{black}{show} the performance at the standard probability threshold of $50\%$ \textcolor{black}{in Supplementary Fig.~\ref{fig:evaluation-fullband}a and we consider a more stringent classification threshold of $90\%$ in Supplementary Fig.~\ref{fig:evaluation-fullband}b}.
For the S/N test, we average over all frequency sub-bands and compare across the 30 S/N bins \textcolor{black}{in Supplementary Fig.~\ref{fig:evaluation-SNR-DR}a}. 

For the drift rate performance evaluation, we simulate ``Single True'' signals between a drift rate range of 0.5 to 10\,Hz/s with a step size of 0.5\,Hz/s. At each drift rate, we loop through the 16 sub-bands each with 1000 samples of randomly selected S/N within the range of 20$-$80. This sums to a total of 16,000 samples per drift rate  \textcolor{black}{and the result can be seen in Supplementary Fig.~\ref{fig:evaluation-SNR-DR}b}.

\begin{table}[ht]
    \centering
    \resizebox{\textwidth}{!}{%
    \begin{tabular}{p{2.3cm}ccccccccc}
\hline
Test type & Source & Freq range & N. Bgs & N. Tests  & S/N range & N$_\mathrm{bin}$ S/N & $\mid$DR$\mid$ &  N$_\mathrm{bin}$ DR\\
 &  & (MHz) &  &  &  & & (Hz/s) & \\
\hline
\multirow{2}{*}{Other models} & HIP\,15638 & 1418.7$-$1475.1 &  4928  & \multirow{2}{*}{24000}  &  \multirow{2}{*}{20$-$80} &  \multirow{2}{*}{30} &  \multirow{2}{*}{0$-$8} &  \multirow{2}{*}{n/a}\\
 & \textcolor{black}{HIP\,13402} & \textcolor{black}{1644.5$-$1700.7} &  \textcolor{black}{4928}  &  & &  &  & \\

Full-band \& & \multirow{2}{*}{HIP\,15638} & \multirow{2}{*}{1023.9$-$1926.3} & \multirow{2}{*}{78848} & \multirow{2}{*}{480000} & \multirow{2}{*}{20$-$80} & \multirow{2}{*}{30} & \multirow{2}{*}{0$-$8} & \multirow{2}{*}{n/a}\\
S/N \\
Drift rate & HIP\,15638 & 1023.9$-$1926.3 & 78848 & 320000 & 20$-$80 & n/a & 0.5$-$10 & 20\\
%Generalized & HIP\,15638 & 1418.7$-$1475.1 &  4928  & 6000 & 20 & 1 & 1$-$2  & n/a\\
Latent Space & HIP\,110750 &  1531.5$-$1587.9 & 4928 & 4928  & n/a & n/a & n/a & n/a\\
\hline
\end{tabular}
}

\caption{Summary of the simulated data parameters for the test benches. We list the test type for each test bench. The source name of the cadence and the frequency range used are shown. The corresponding number of unique background snippets (N. Bgs) is listed, from which we randomly draw the said number of test samples (N. Tests) to inject simulated signals. The S/N range for each case is shown as well as the number of S/N bins evenly distributed within the range --- except when the S/N are randomly chosen from the range specified we denote the cell with ``n/a''. Similarly, we list the absolute value of drift rate (DR) and its number of bins. For the latent space test, no simulated signal is injected. }
\label{tab:background-testbench}
\end{table}

\subsection*{Alternative Model Hyperparameter Tuning \label{sec:hyperparameter}
}
\textcolor{black}{
Four alternative ML models were considered for this SETI project, namely Random Forest, SVM, ANN, and CNN. 
We conduct a large-scale hyperparameter search in order to find the best configurations for each model. 
We search 1000 different variations for Random Forest and SVM. For the ANN and CNN, we search 3000 variations and each time train with 100 epochs. 
Using a Bayesian optimization technique, we sample a total of 128 variations out of all these possibilities, executed on a $k$-fold cross validation set with 5 splits. In other words, a total of 640 variations were trained. Parameters that have led to the best performing models are highlighted in bold in Table~\ref{tab:4model_param}. 
The averaged and best AUC from this search can be found in Table~\ref{tab:4model_AUC}. 
The averaged AUCs all have very small standard deviations, which is to say that the performance of each of these four models are pretty consistent across the hyperparameter space searched, and only marginal improvement has been achieved by fine tuning the hyperparameters over the default, `out-of-the-box' configurations.
Note that these AUCs are results of the hyperparameter search using the validation data set, which are not the same as the AUCs listed in Table~\ref{tab:f1} from the test data. The AUCs from the hyperparameter search appear to be a little higher than those from the test data. This is expected as some degree of overfitting can happen when optimizing on the validation set. 
}

\begin{table}[ht]
    \centering
        \resizebox{.55\textwidth}{!}{%
   { \color{black}\begin{tabular}{p{5cm}c}
   \Xhline{2\arrayrulewidth}
\multicolumn{2}{c}{Random forest }\\
\Xhline{2\arrayrulewidth}
Parameters &  Values\\
\hline
Number of estimators & 10, 50, 100, 200, 500, \textbf{1000} \\
Parameter criterion & gini, \textbf{entropy} \\
Maximum depth & 5, 10, \textbf{20}, 30, None\\
Maximum features & auto, \textbf{sqrt}, log2\\
\Xhline{2\arrayrulewidth}
\multicolumn{2}{c}{SVM}\\
\Xhline{2\arrayrulewidth}
Parameters &  Values\\
\hline  
C regularization & 0, \(1\times 10^{-6}\), \textbf{1}, 2, 10, 50, 100\\
Degree & 1, 2, 3, 5, \textbf{7}, 9\\
Kernels & linear, poly, rbf, \textbf{sigmoid}\\
\Xhline{2\arrayrulewidth}
\multicolumn{2}{c}{Artificial neural net} \\
\Xhline{2\arrayrulewidth}
Parameters &  Values\\
\hline
First layer weights & 256, \textbf{512}  \\
Number of first layers  & \textbf{1}, 2  \\
Activations for the first layers & \textbf{ReLU}, sigmoid \\
Second layer weights &  128, \textbf{256} \\
Number of second layers  & 1, \textbf{2} \\
Activations for the second layers &  \textbf{ReLU}, sigmoid \\
Third layer weights & \textbf{32}, 64 \\
Number of third layers  & \textbf{1}, 2  \\
Activations for the third layers &  \textbf{ReLU}, sigmoid \\
Final activation &  \textbf{sigmoid}, softmax \\
Learning rate  & \textbf{0.001}, 0.0001, 0.00001 \\
\Xhline{2\arrayrulewidth}
\multicolumn{2}{c}{Convolutional neural net} \\
\Xhline{2\arrayrulewidth}
Parameters &  Values\\
\hline
First-layer filter size & 8, \textbf{16}, 24 \\
Number of first layers  & 1, 2, \textbf{3}  \\
Activations for the first layers & \textbf{ReLU}, sigmoid \\
Second-layer filter size & \textbf{32}, 64, 128  \\
Number of second layers  & 1, 2, \textbf{3}  \\
Activations for the second layers & \textbf{ReLU}, sigmoid \\
Third-layer filter size &  \textbf{64}, 128 \\
Number of third layers  & 1, 2, \textbf{3} \\
Activations for the third layers &  \textbf{ReLU}, sigmoid \\
Hidden layer weights & \textbf{256}, 128 \\
Final activation &  \textbf{sigmoid}, softmax \\
Learning rate  & \textbf{0.001}, 0.0001, 0.00001 \\
\hline

\hline
\end{tabular}}
}
\caption{ \textcolor{black}{Parameter values searched for each of the four sets of alternative models. The values in bold indicate the best performing model.}}
\label{tab:4model_param}
\end{table}

\begin{table}[ht]
    \centering
        \resizebox{.6\textwidth}{!}{%
    \begin{tabular}{cccc}
\hline
Model & \textcolor{black}{N. variations} & \textcolor{black}{Averaged AUC} & \textcolor{black}{Best AUC}   \\
\hline
Random Forest & \textcolor{black}{1000} & \textcolor{black}{0.77(1)}  & \textcolor{black}{0.79} \\
SVM & \textcolor{black}{1000} & \textcolor{black}{0.72(9)} & \textcolor{black}{0.78} \\
ANN  & \textcolor{black}{3000} & \textcolor{black}{0.73(1)} & \textcolor{black}{0.74}\\
CNN  & \textcolor{black}{3000} & \textcolor{black}{0.95(2)} & \textcolor{black}{0.98}\\
\hline
\end{tabular}
}
\caption{ \textcolor{black}{The number of variations of parameter values searched, the averaged AUCs with standard deviation shown in brackets, as well as the best AUCs of the four alternative ML models from the hyperparameter tuning using the validation data set.}}
\label{tab:4model_AUC}
\end{table}

\subsection*{Further details on the Performance Assessments} \label{sec:threshold}
\begin{figure}[H]
\begin{subfigure}{.5\textwidth}
  \centering
  \includegraphics[width=\linewidth]{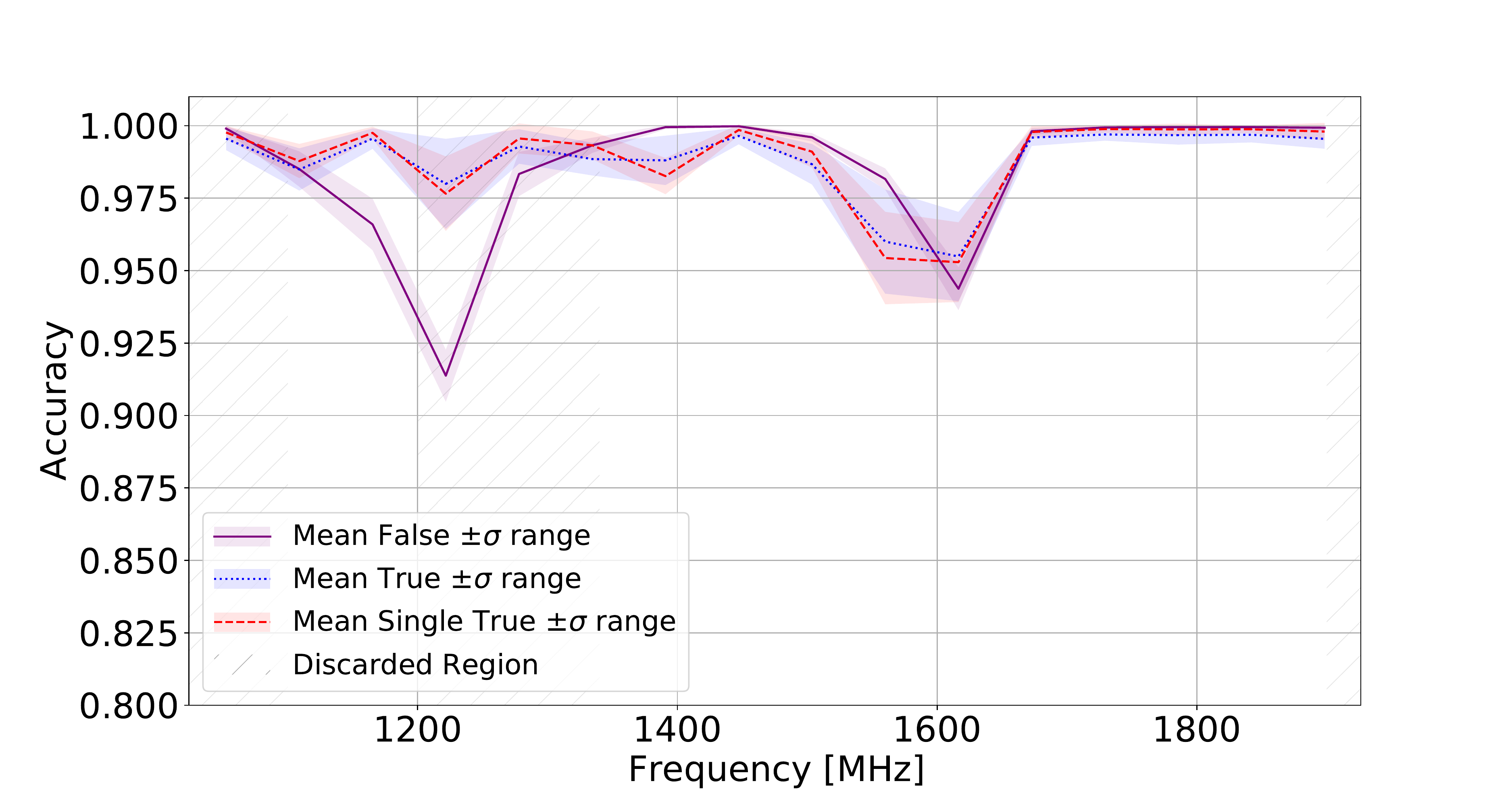}
\end{subfigure}%
\begin{subfigure}{.5\textwidth}
  \centering
  \includegraphics[width=\linewidth]{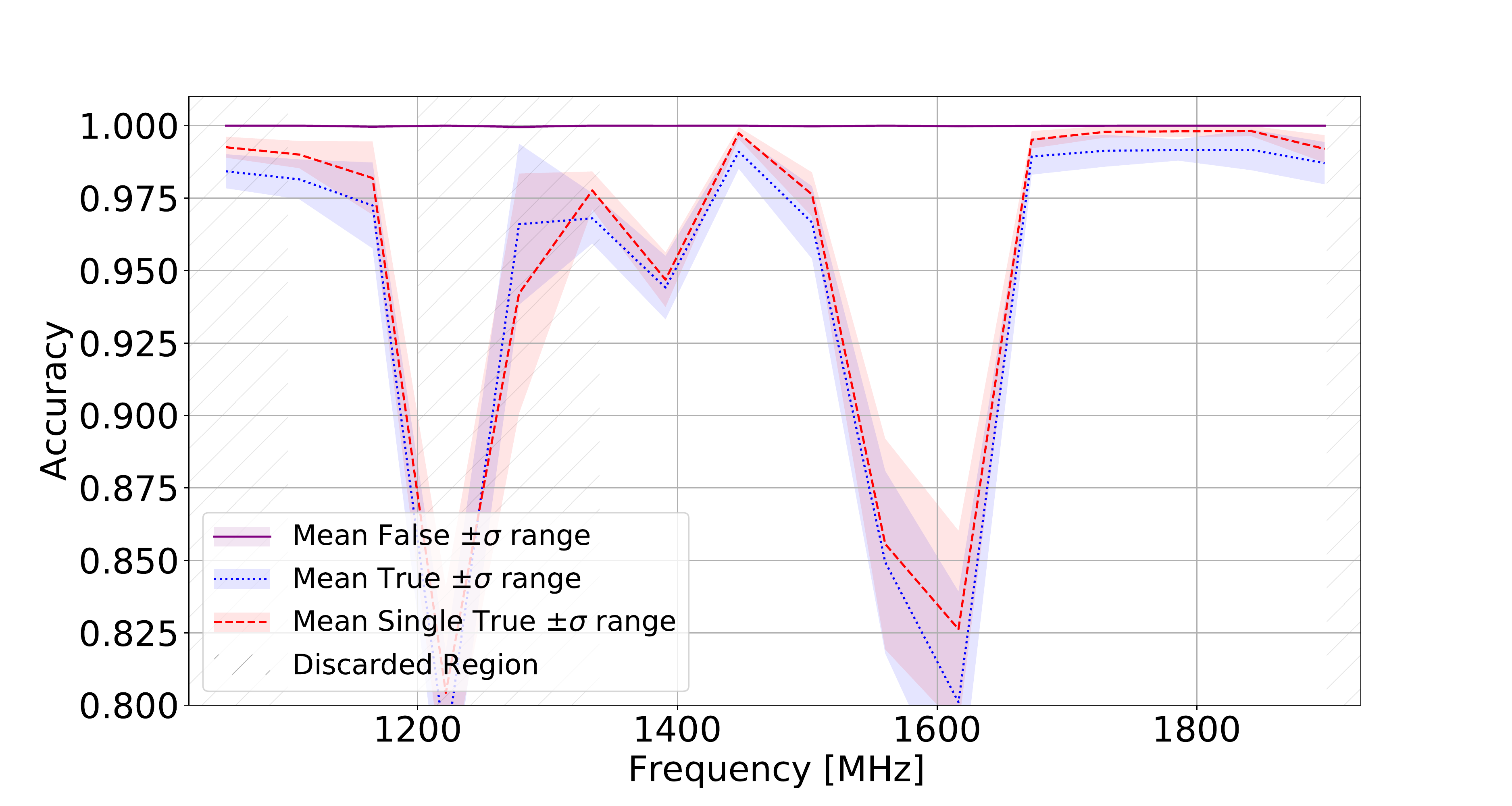}
\end{subfigure}
\caption{Performance evaluation across the full band for a classification threshold of (a) 50\% and (b) 90\% as a function of the frequency band. The regions of the band discarded due to instrumental artifacts are shaded in gray. The colored bands represent one standard deviation.}
\label{fig:evaluation-fullband}
\end{figure}

\begin{figure}[H]
\begin{subfigure}{0.5\textwidth}
  \centering
  \includegraphics[width=\linewidth]{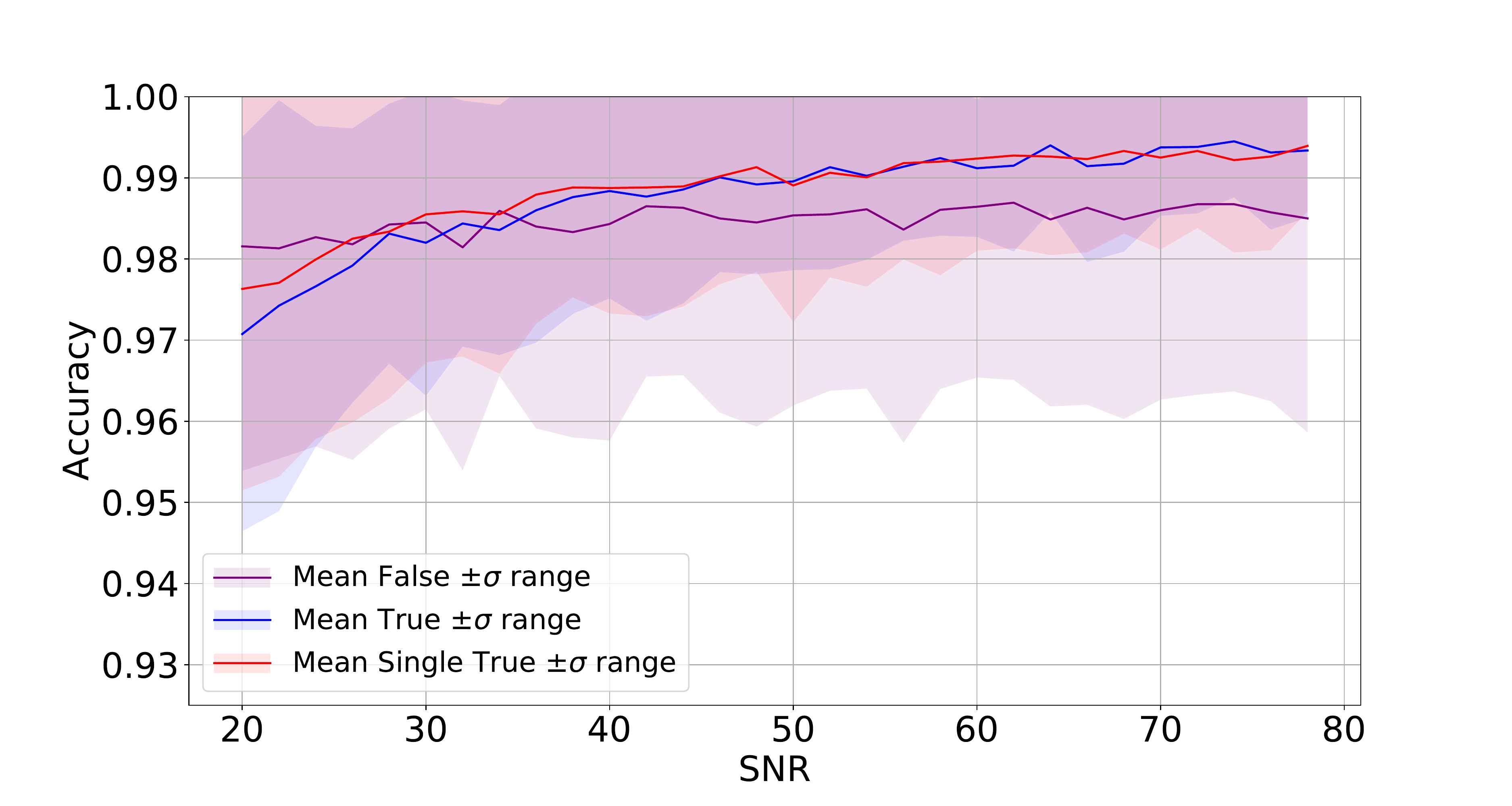}
  \end{subfigure}%
\begin{subfigure}{0.5\textwidth}
  \centering
  \includegraphics[width=\linewidth]{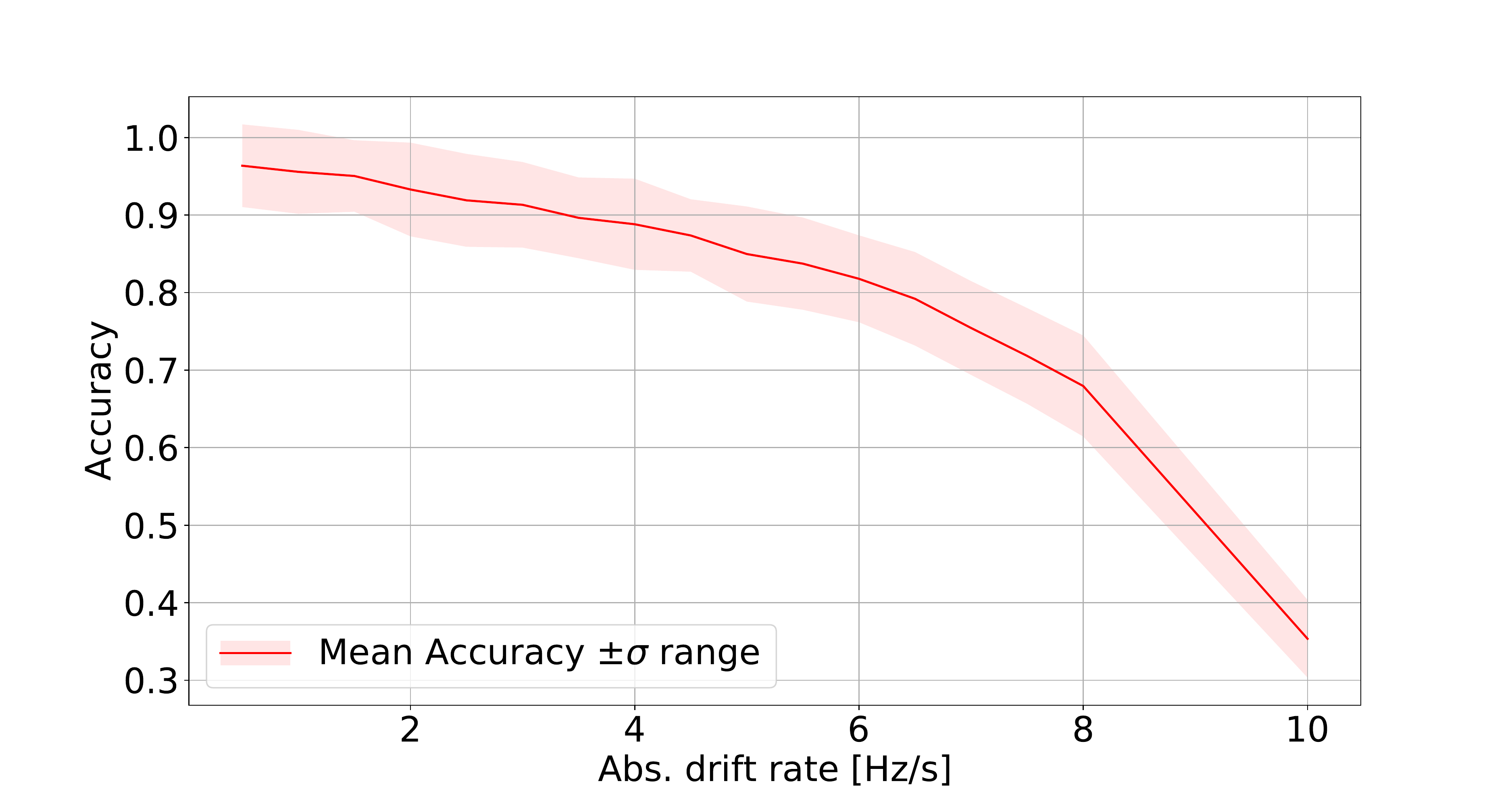}
    \end{subfigure}%
\caption{(left) Performance evaluation as a function of S/N for a threshold of 50\%. The colored bands represent one standard deviation. (right) Performance evaluation on ``Single True'' signals as a function of drift rate for a threshold of 90\%. The colored bands represent one standard deviation.}
\label{fig:evaluation-SNR-DR}
\end{figure}

\textcolor{black}{
For the benchmark test against other ML models, we compute the standard metrics of precision, recall and F1 scores as recorded in Supplementary Table~\ref{tab:f1}.
Precision describes how well the positive labels are determined, and is defined by the number of true positives (TP) divided by the number of true positives plus false positives (TP+FP). 
Recall is defined as TP divided by the total positive samples (true positives plus false negatives; TP+FN). 
The F1 score is a measure of both of those metrics, where F1=2$\times$(Recall $\times$ Precision) / (Recall + Precision).}
\textcolor{black}{Supplementary  Figs.~\ref{fig:evaluation-fullband} and~\ref{fig:evaluation-SNR-DR} show our model accuracy as a function of the observing frequency, S/N and drift rate.}

\begin{table}[ht]
    \centering
        \resizebox{.55\textwidth}{!}{%
    \begin{tabular}{ccccc}
\hline
Model  & \textcolor{black}{Best AUC}  & Precision & Recall & F1 \\
\hline
Random Forest &  \textcolor{black}{0.7123} & 0.641 &  0.956 & 0.767\\
SVM &  \textcolor{black}{0.7115}  & 0.643 & 0.951 & 0.767\\
ANN  &  \textcolor{black}{0.6768} & 0.636 & 0.659 & 0.647\\
CNN  &  \textcolor{black}{0.9817} & 0.992 & 0.934 & 0.959\\
Our Model & \textcolor{black}{0.9993} & 0.994 & 0.985 & 0.989 \\
\hline
\end{tabular}
}
\caption{ \textcolor{black}{The best AUCs,} precision, recall, and F1 scores of the five ML models applied on the test data set.}
\label{tab:f1}
\end{table}

\subsection*{Latent Space Analysis} \label{sec:latentspace}
\begin{figure}[H]
\centering
\includegraphics[scale=0.5]{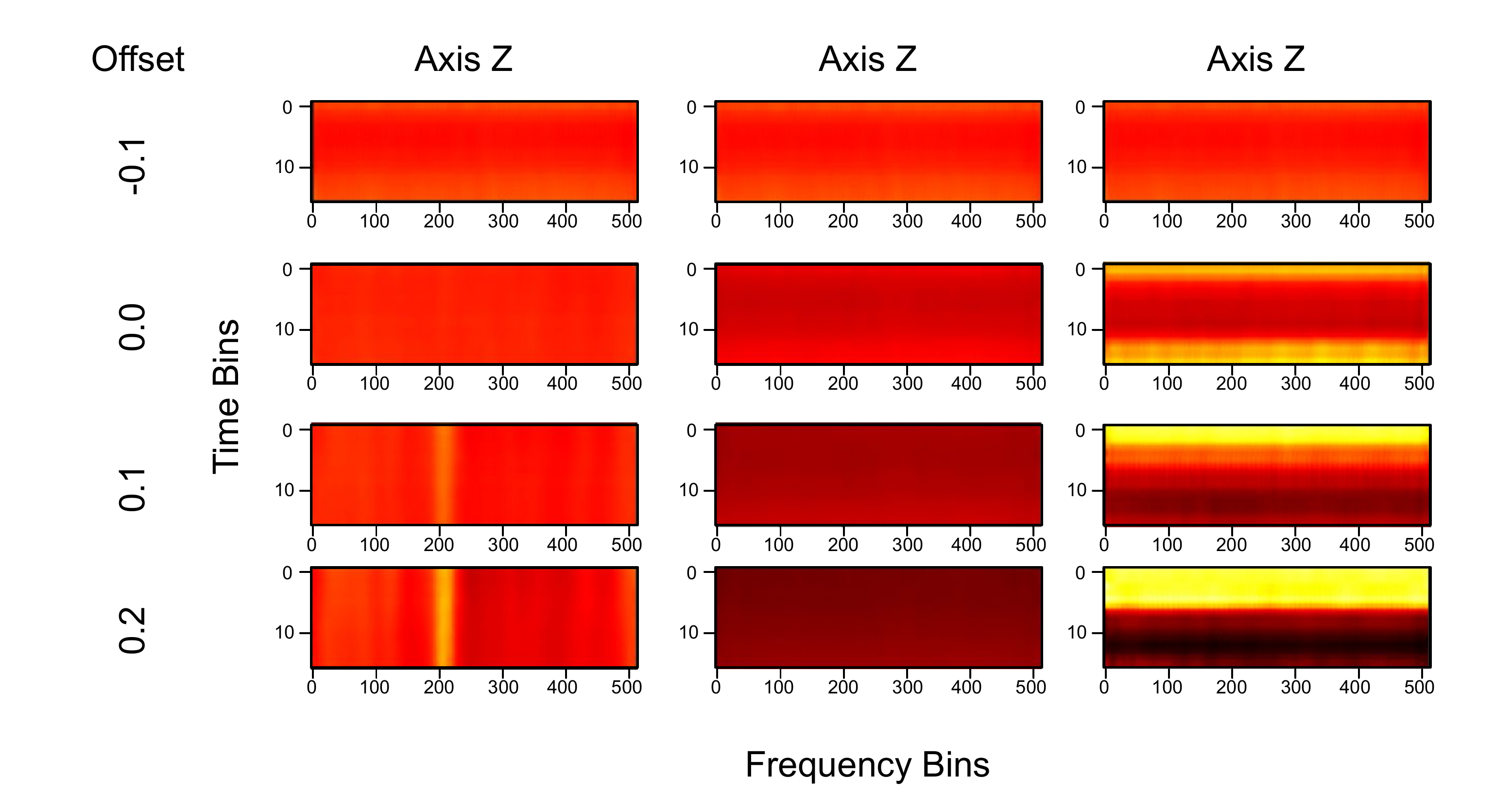}

\caption{Spectrograms showing the effects of varying the first three axes of the latent space vectors. The perturbations are in steps of 0.1 and are indicated on the left hand side of the plots. }
 \label{fig:latentspace3axes}
\end{figure}

In order to understand and interpret how our ML algorithm makes classification decisions, we study the latent space of the decoder with the goal of finding out whether the network is able to disentangle each of the learned features.
The first three axes appear to produce human interpretable results 
and so we further study their feature vectors and we perturb each axis in steps of 0.1. These variations are then fed back into the decoder and we visualize the resultant spectrograms in \textcolor{black}{Supplementary} Fig.~\ref{fig:latentspace3axes}.
Axis $z$1 looks to have encoded the feature of S/N as perturbing this axis creates a brighter signal. Axis $z$2 appears to represent the level of background noise and axis $z$3 corresponds to instrumental artifacts which are commonly seen as bright horizontal bands across part of the spectrum. We interpret this outcome as evidence that the network has learnt relevant features from the dataset, helping it make the right decisions during the classification stage. 

\textcolor{black}{To quantify this performance, we implement a disentanglement metric \cite{bvae} by simulating idealized signals 
with four axes of latent space vectors, namely drift rate, width of the signal, SNR and frequency. 
We then applying a linear classifier on the latent space vectors to differentiate between different perturbations of simulation parameters. If disentanglement is successful, a linear classifier is sufficient in separating changes in the four axes just from the latent vectors. The accuracy of the benchmark \(\beta\)-VAE from \cite{bvae} achieved a maximum accuracy of 99.2\%  where as the traditional VAE had only 61.6\%. Our model returns an accuracy of 70.1\%, which is superior compared to that of a traditional VAE model and confirming our improvement in disentanglement and interpretability scores. The drop in disentanglement score in respect to the \(\beta\)-VAE is due to the fact that our model used a smaller \(\beta =1.5\) factor rather than \(\beta=4\) factor from the original implementation. This difference in factors was to balance the custom loss function we constructed for the cadence clustering.}

\subsection*{Search Pipeline} \label{sec:pipeline}
\begin{figure}[H]
\includegraphics[scale=0.26]{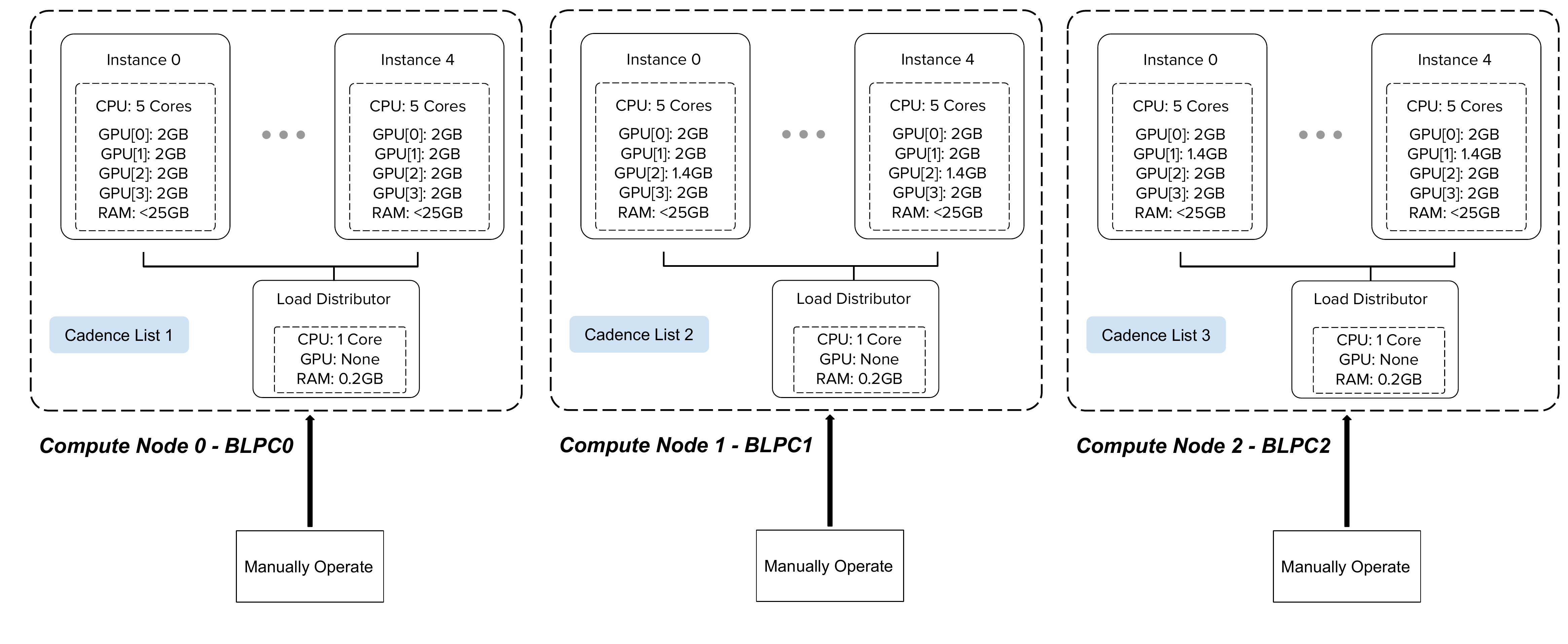}
\caption{Figure depicts the orchestration of singularity containers and how the computing resources were distributed to perform individual searches.} \label{fig:computenode}
\end{figure}

We develop a processing pipeline in order to systematically analyze the 120\,TB of GBT 1.1--1.9\,GHz data. 
Within each 4096-channel snippet, each search is completely independent and hence can be easily parallelized. 
We have access to three compute nodes at the Berkeley data centre for this analysis (see \textcolor{black}{Supplementary} Table~\ref{tab:hw} for the hardware specifications), therefore we divide up the dataset into three equal sub-sets to run on the three machines independently as shown in \textcolor{black}{Supplementary} Fig.~\ref{fig:computenode}. On each machine, we launch five parallel instances of the processing that independently work on different cadences. The status of the processing on each machine is monitored and managed manually.

Initially, we divide up the full frequency band into 16 sub-bands. This allows us to load observations faster into memory compared to reading in and processing each 4096-channel snippet one by one, because of the switching cost during input/output (I/O) vs processing. The I/O was done using the python package  \textsc{Blimpy}\footnote{https://github.com/UCBerkeleySETI/blimpy} as described in \cite{blimpy}. 
We concatenate the six observations of a cadence before feeding it into the neural network, which is built using the Tensorflow API and is GPU-accelerated. Furthermore we distribute the training across multiple GPUs implementing a typical asynchronous mirrored approach. Because we mirror each model on different GPUs we can afford to execute the model on batch sizes of $10^4$ snippets given the GPU memory constraints to further parallelize execution. Finally the features are fed into the Random Forest classifier which runs in parallel on the CPU. 

%Times
The search pipeline, disregarding I/O, takes $<$5\,s to search $1/16^{\mathrm{th}}$ of the band. The I/O and the preprocessing take the most amount of compute time despite it being implemented to compute in parallel and with just-in-time compilers (JIT) such as \textsc{NUMBA}\footnote{https://numba.pydata.org/}, taking $50 - 60$\,s to run on the same $1/16^{\mathrm{th}}$ band. The search time on a single 30\,min cadence is $18.7 \pm 1.6$\,min. 
The variation is mostly due to cache preloading for some data. 
 However the runtime of the core of the search pipeline disregarding the I/O stays relatively constant from cadence to cadence.

%further speed up
We achieve further speed-up via a custom container orchestration using \textsc{Singularity}\footnote{https://sylabs.io/} and the schematic is shown in \textcolor{black}{Supplementary} Fig.~\ref{fig:computenode}. By running multiple instances of the same search on smaller subsystems on the same compute node, it scales faster in practice than running a single instance on all the resources \cite{singularity}. This is because each search is completely independent from each other and hence does not require sharing memory or data. This is far more efficient than parallelizing individual operations such as I/O where data needs to then be recombined together.

\begin{table}[ht]
    \centering
    \begin{tabular}{llll}
\hline
 & Compute Node 0 & Compute Node 1 & Compute Node 2\\
\hline
CPU   & Intel Xeon E5-2630  & Intel Xeon E5-2630&	Intel Xeon Silver 4210 	\\
GPU[0] & NVIDIA TITAN X	    & NVIDIA TITAN X	&	NVIDIA TITAN X	\\
GPU[1] & NVIDIA TITAN X	    & NVIDIA TITAN XP	&	NVIDIA GTX 1080	\\
GPU[2] & NVIDIA TITAN X	    & NVIDIA GTX 1080	&	NVIDIA GTX 1080 Ti	\\
GPU[3] & NVIDIA TITAN X	    & NVIDIA TITAN XP	&	NVIDIA GTX 1080 Ti	\\
RAM   & 256\,GB	            & 256\,GB	            &	196\,GB	\\

\hline
\end{tabular}
\caption{The hardware specifications for each compute node used in this analysis. }
\label{tab:hw}
\end{table}

\subsection*{Signal of interest visualization} \label{sec:visualization}
\begin{figure}[H]
\begin{subfigure}{.5\textwidth}
  \centering
  \includegraphics[width=0.95\linewidth]{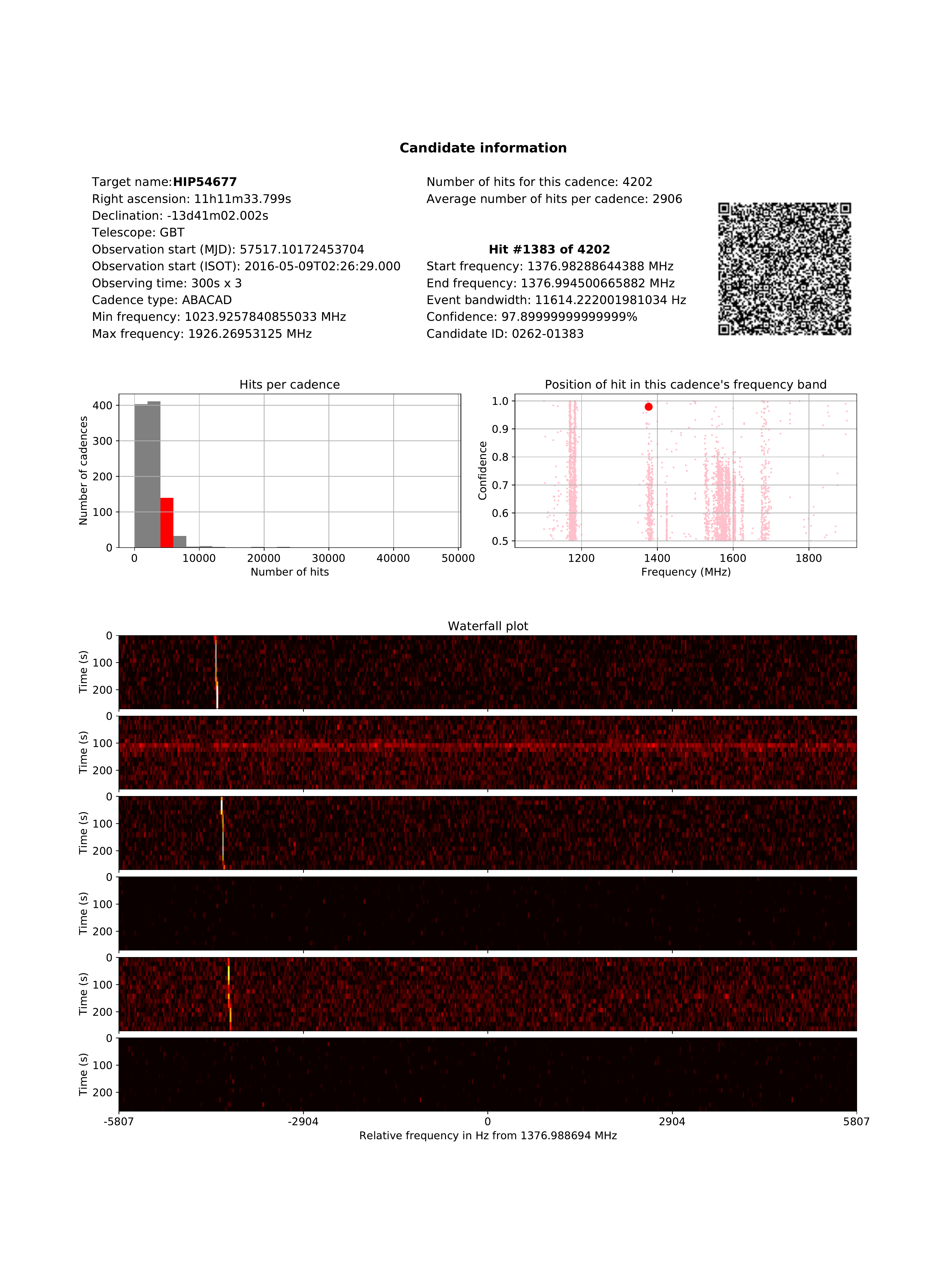}
\end{subfigure}%
\begin{subfigure}{.5\textwidth}
  \centering
  \includegraphics[width=0.95\linewidth]{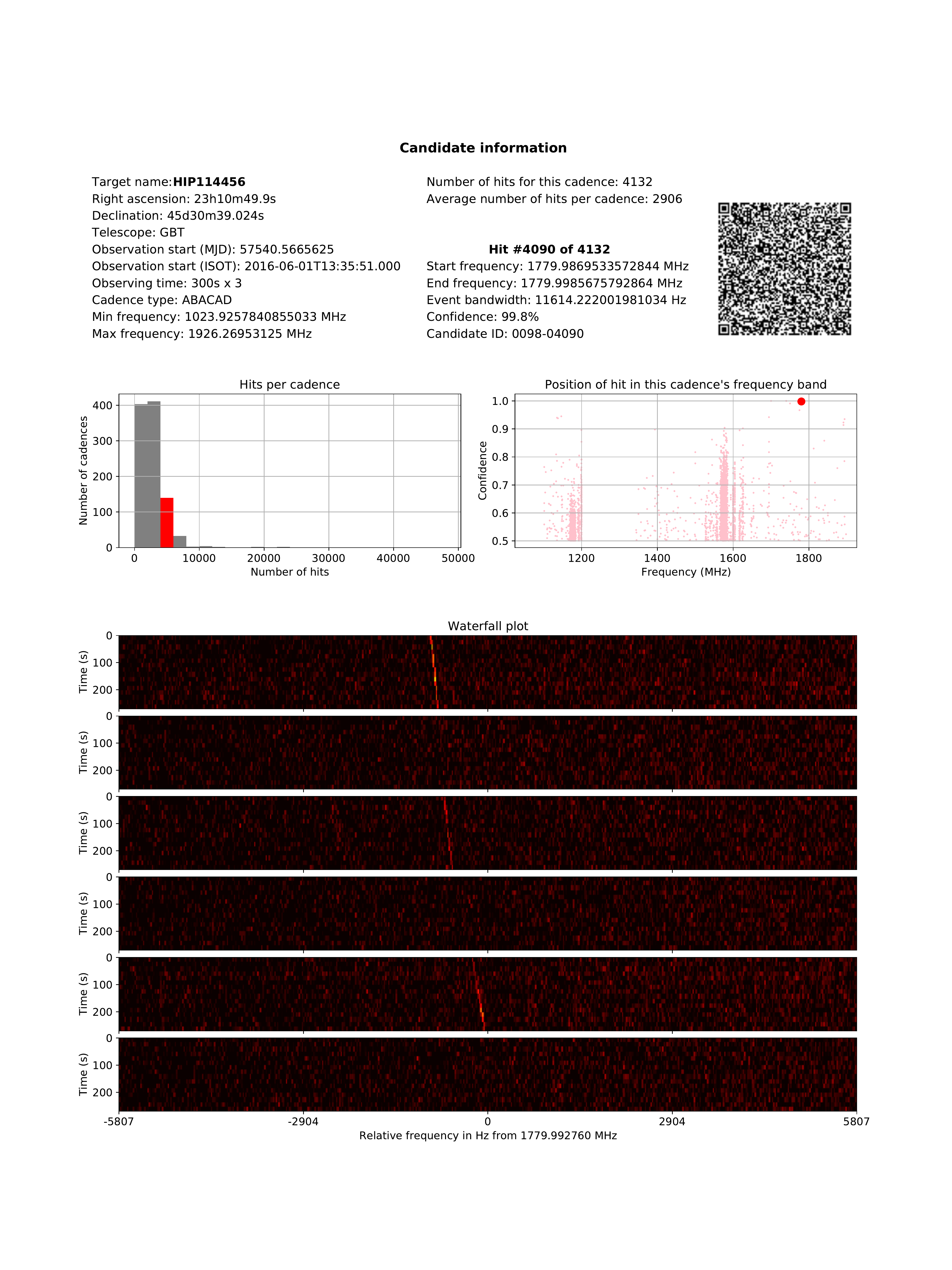}
\end{subfigure}
\caption{Two examples of the diagnostic plots. (Left) One of the top $\NTOPCAND$ \textcolor{black}{signals of interest} described in Table~\ref{tab:top10}. (Right) An interesting event returned by the ML model that is rejected as \textcolor{black}{a} SETI candidate upon human inspection because of the non-uniform drift rate across the three ON-scans.}
\label{fig:2cands}
\end{figure}

In order to display the information for each \textcolor{black}{signal of interest} snippet and to visually assess them, we create diagnostic plots for each of the 20,515 events returned by the ML model using a Python script.
\textcolor{black}{Supplementary} Fig.~\ref{fig:2cands} shows two examples of these diagrams, one for an event measured in the observation of HIP\,54677 (described in Table~\ref{tab:top10} as MLc5, one of the top $\NTOPCAND$ \textcolor{black}{signals of interest} of our search) and one for an event in HIP\,114456, which is ultimately rejected upon human assessment due to the non-uniform drift rate across the three ON observations.

%Header text
The header contains information about both the whole cadence and the \textcolor{black}{signal of interest} snippet from that cadence returned by the ML model. The information on the cadence includes the catalog name of its target star (the ON observation), its celestial coordinates (right ascension and declination), the telescope used, the start time of the observation written in both modified Julian day (MJD) and in ISO~8601 format for date and time (ISOT), the time spent observing ON-target, the cadence type (ABABAB or ABACAD), as well as the minimum and maximum frequency of this cadence recorded during the observation. Adding to this information are the number of events identified by the ML model for this specific cadence and the average number of events identified for all cadences. 
The information presented on the \textcolor{black}{signal of interest} includes the start and end frequencies of the snippet, the bandwidth of the snippet as well as the confidence rating according to the ML model. Each \textcolor{black}{signal of interest} is given a numerical identification composed of a number representing the cadence and a number representing the event snippet in that cadence. The header is juxtaposed with a QR code which encodes the ID of the  \textcolor{black}{signal of interest}, the cadence target name, the start and end frequencies of the snippet as well as paths to the visualizer itself and to each of the six HDF5 files that make up the cadence. Paths correspond to locations on the Breakthrough Listen computers. 

\
%Middle two plots and waterfall
Two analytical plots are included in the middle of the diagram. The plot on the left is a histogram which classifies the cadences by the number of events identified in them, with the bin of the cadence in question highlighted in red. This is used to give a sense whether the specific cadence contains an \textcolor{black}{un}usually high amount of events (\textcolor{black}{signals of interest}) or not, as that might indicate an observation that is heavily contaminated by RFI and thus potentially less reliable.
The plot on the right shows the positions of all events of a cadence in frequency space (i.e. where they fall in the bandwidth of the cadence) as well as the confidence ratings of the \textcolor{black}{signals of interest}. The \textcolor{black}{signal of interest} in question is represented by a large red dot in this plot. This plot helps to assess whether the specific \textcolor{black}{signal of interest} comes from a region of the frequency band that has a large number of events (higher chance that the \textcolor{black}{signal of interest} is also RFI) or if the region is relatively empty of events (higher chance that the \textcolor{black}{signal of interest} is genuinely special).
Finally, the bottom half of the diagram is dedicated to the waterfall plot, a representation of frequency vs. time showing the \textcolor{black}{signal of interest} snippet across each of the six ON and OFF observations with lighter color signifying higher intensity. The waterfall plot is created using a modified version of the Python package \textsc{blimpy}\footnote{https://github.com/UCBerkeleySETI/blimpy} and is downsampled by a factor of 8 in frequency, the same as the input dimension to the ML model. In other words, each frequency bin in the plot has a resolution about $\sim 22.4$\,Hz.

All 20,515 diagrams generated are collated into an mp4 file to be viewed as a movie with four frames per second. The full video can be found at this weblink\footnote{https://www.youtube.com/watch?v=iSdVfOwPVCI}.

\subsection*{Comparison with TurboSETI}
One of the most extensively used SETI algorithms in Breakthrough Listen's analysis is \textsc{turboSETI}\footnote{https://github.com/UCBerkeleySETI/turbo\_seti} \textcolor{black}{as described in} \cite{enriquez2017turbo,turboseti}, which is designed to detect these narrowband drifting signals by implementing the ``tree de-Doppler'' algorithm for incoherent Doppler acceleration searches\cite{Siemion2013}. 
\textcolor{black}{One of the reasons} we have chosen to work with this \textcolor{black}{GBT} dataset is because a thorough ETI search has previously been conducted on it using a standard de-Doppler technique \cite{enriquez2017turbo,Price2020}. This provides a way to benchmark our algorithms as we can compare the output of two completely different methods. 
Out of the $\NCADENCES$ cadences analyzed here, 688 were studied by \cite{enriquez2017turbo} and 755 were in the sample used by \cite{Price2020}. The small mismatch in the source list is because we have rejected several cadences that have incomplete observations with less than 16 time bins, which are incompatible with our ML algorithm. 

\begin{figure}[H]
\centering
\includegraphics[scale=0.4]{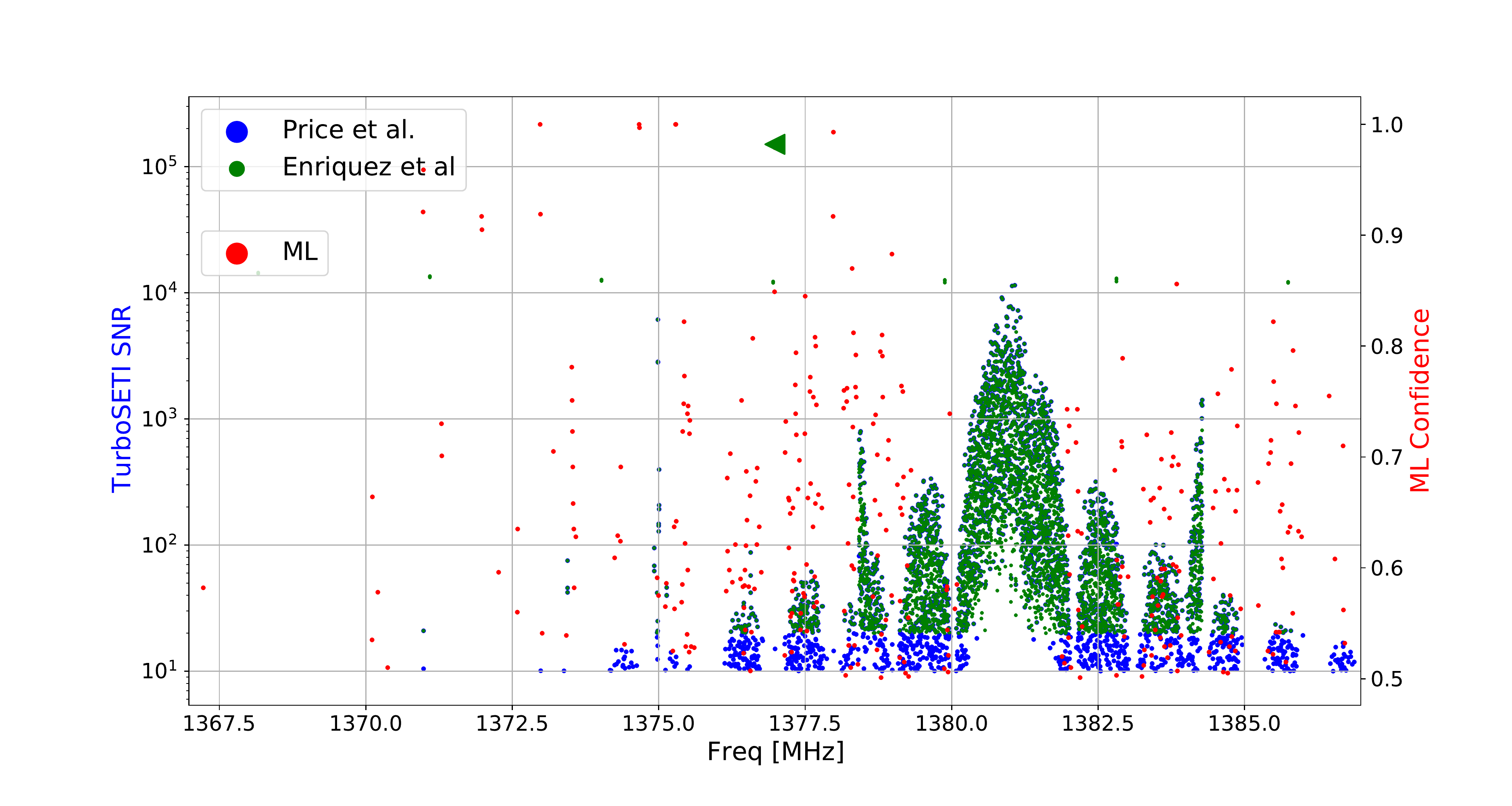}
\caption{The S/N-confidence vs frequency distribution of the hits identified by \textsc{turboSETI} as reported by \cite{Price2020} (blue) and by \cite{enriquez2017turbo} versus events detected by our ML model (red) in the cadence of target HIP\,54677. The ML event that corresponds to MLc5 is marker by a black triangle symbol.} \label{fig:ML-Turbo}
\end{figure}

\textcolor{black}{Supplementary} Fig.~\ref{fig:ML-Turbo} \textcolor{black}{shows} a comparison of the hits (not yet grouped per cadence to be considered as events) reported by these two \textsc{turboSETI} searches versus the events found by our ML algorithm on target HIP\,54677 in the frequency ranges where we have detected our ETI \textcolor{black}{signal of interest} MLc5. The region around the \textcolor{black}{signal of interest} is clearly empty of any other detections.
The blue and green TurboSETI hits essentially outline the GPS signal in the observing band. In fact eight out of the 11 prime signals-of-interest presented in \cite{enriquez2017turbo} are from that particular region of 1370$-$1380\,MHz. \textcolor{black}{Overall, we} find that for every cadence, on average 64\% of the events (with a deviation of 15\%) flagged by our ML were not found by \cite{enriquez2017turbo} and on average 61\% (with a deviation of 37\%) were not found by \cite{Price2020}. It thus appears that the events identified by the two search \textcolor{black}{algorithms} are quite distinct from each other. 

Comparing our \textcolor{black}{method to the conventional} \textsc{turboSETI}, we find that our ML mode is successful in returning fewer false positives and more convincing \textcolor{black}{signals of interest}. 
Part of this is due to the fact that our ML model is built to consider the cadence pattern as a whole, whereas \textsc{turboSETI} searches each observation separately to produce hits, which then require a secondary grouping stage to cluster hits that come from the same \textcolor{black}{signal of interest}. 
We also note that \textcolor{black}{our search has been able to cover a wider drift rate range up to $\pm$10\,Hz/s} and none of our top $\NTOPCAND$ \textcolor{black}{signals of interest} were identified \textcolor{black}{by the \textsc{turboSETI} searches} \cite{enriquez2017turbo,Price2020}.
Our ML algorithm also has some disadvantages compared to \textsc{turboSETI}. One of them is that unlike \textsc{turboSETI}, we do not directly obtain drift rate and S/N as output of the pipeline. We also do not know precisely where the signal of interest lies in terms of observing frequency and only know within which 4096-channel snippet ($\sim$11.6\,kHz width) it is.

\subsection*{Code Availability}
The code is available for review here at (\url{https://github.com/PetchMa/ML_GBT_SETI})

\subsection*{Data Availability}
All data used in this manuscript are stored as high-resolution \textsc{filterbank} and \textsc{HDF5} format collected and generated from observations by the Robert C. Byrd Green Bank Telescope,  which are available through the Breakthrough Listen Open Data Archive at http://seti.berkeley.edu/opendata. Correspondence and requests for other materials should be addressed to P.M.

\subsection*{Acknowledgements}
Breakthrough Listen is managed by the Breakthrough Initiatives, sponsored by the Breakthrough Prize Foundation. (\url{http://www.breakthroughinitiatives.org}) We are grateful to the staff of the Green Bank Observatory for their help with installation and commissioning of the Breakthrough Listen backend instrument and extensive support during Breakthrough Listen observations. P.M. was supported by the Laidlaw foundation which has funded this project as part of the undergraduate research and leadership funding initiative. S.Z.S. acknowledges that this material is based upon work supported by the National Science Foundation MPS-Ascend Postdoctoral Research Fellowship under Grant No. 2138147. We thank Yuhong Chen for his helpful discussion on the Machine Learning framework. P.M. would like to thank the kind support of Dr. Laurance Doyle and Dr. Sarah Marzen for their generous guidance and encouragement to him when he first began his research career.  

\subsection*{Competing Interests}
The authors declare no competing interests.

\subsection*{Supplementary information}
Diagnostic plots for the remaining seven \textcolor{black}{signals of interest} listed in Table~\ref{tab:top10}. 

\begin{figure}[H]
\begin{subfigure}{.5\textwidth}
  \centering
  \includegraphics[width=0.95\linewidth]{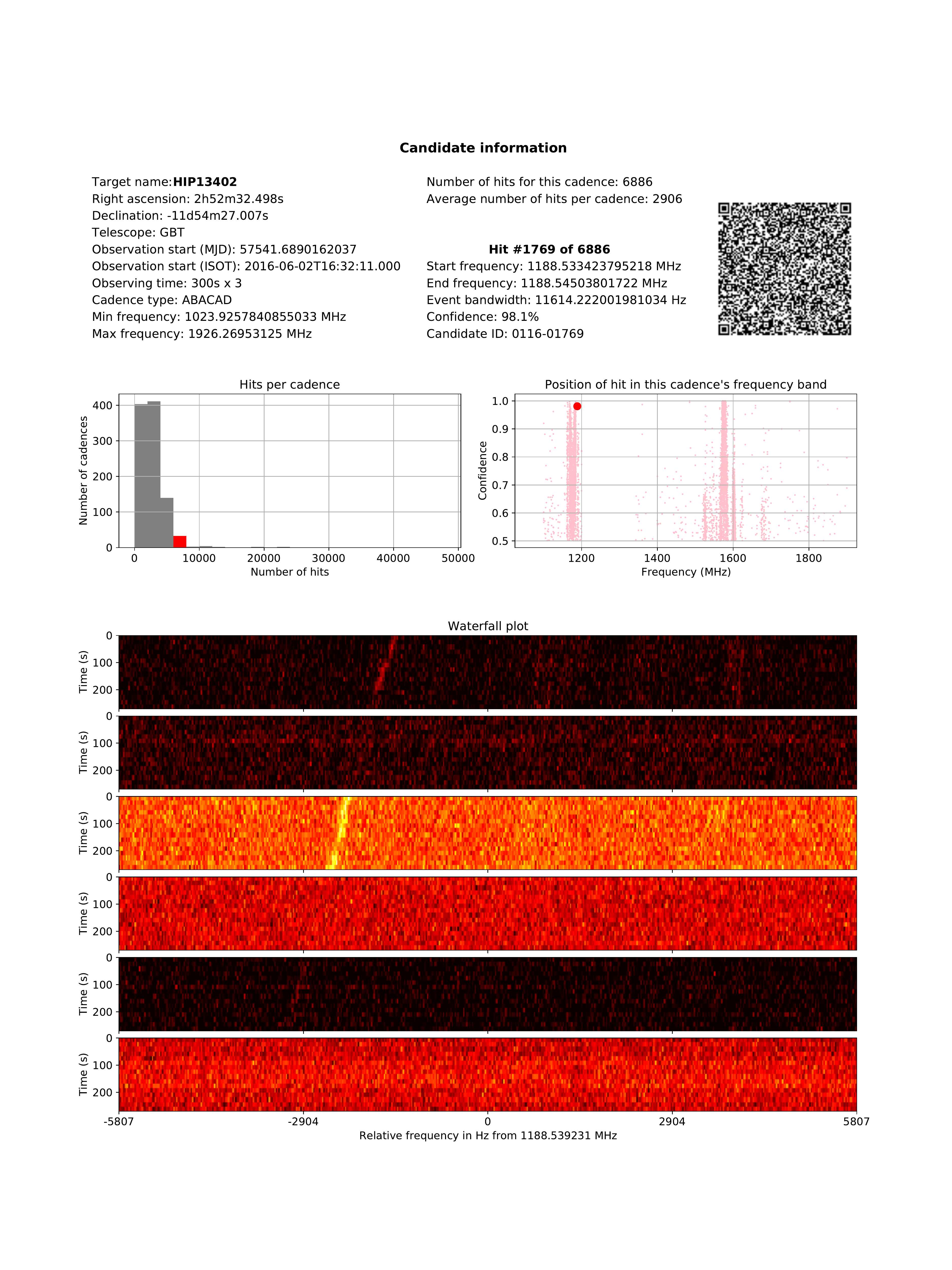}
\end{subfigure}%
\begin{subfigure}{.5\textwidth}
  \centering
  \includegraphics[width=0.95\linewidth]{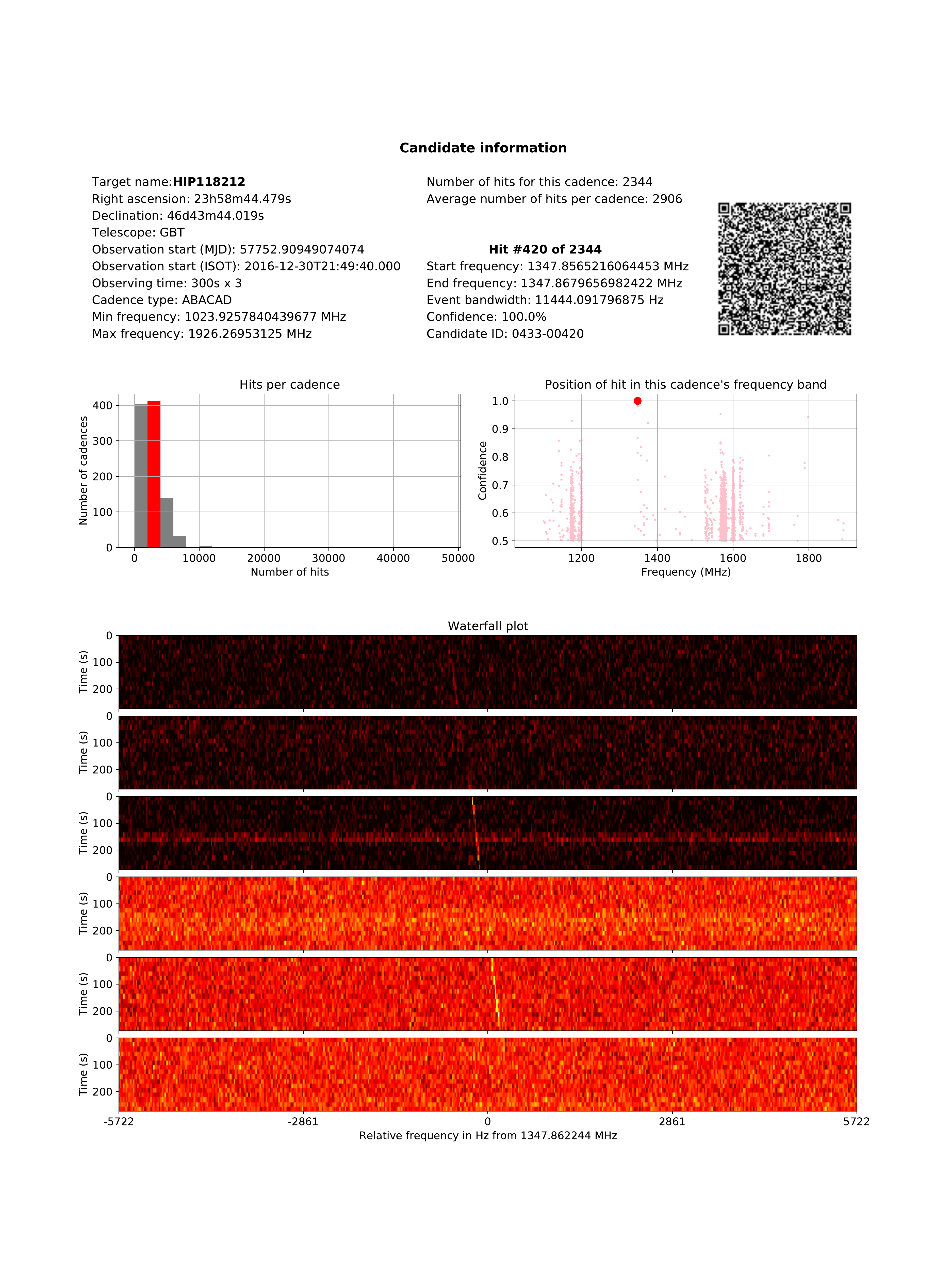}
\end{subfigure}
\caption{Diagnostic plots for (Left) MLc1 (Right) MLc2.}
\label{fig:MLc1-MLc2}
\end{figure}

\begin{figure}[H]
\begin{subfigure}{.5\textwidth}
  \centering
  \includegraphics[width=0.95\linewidth]{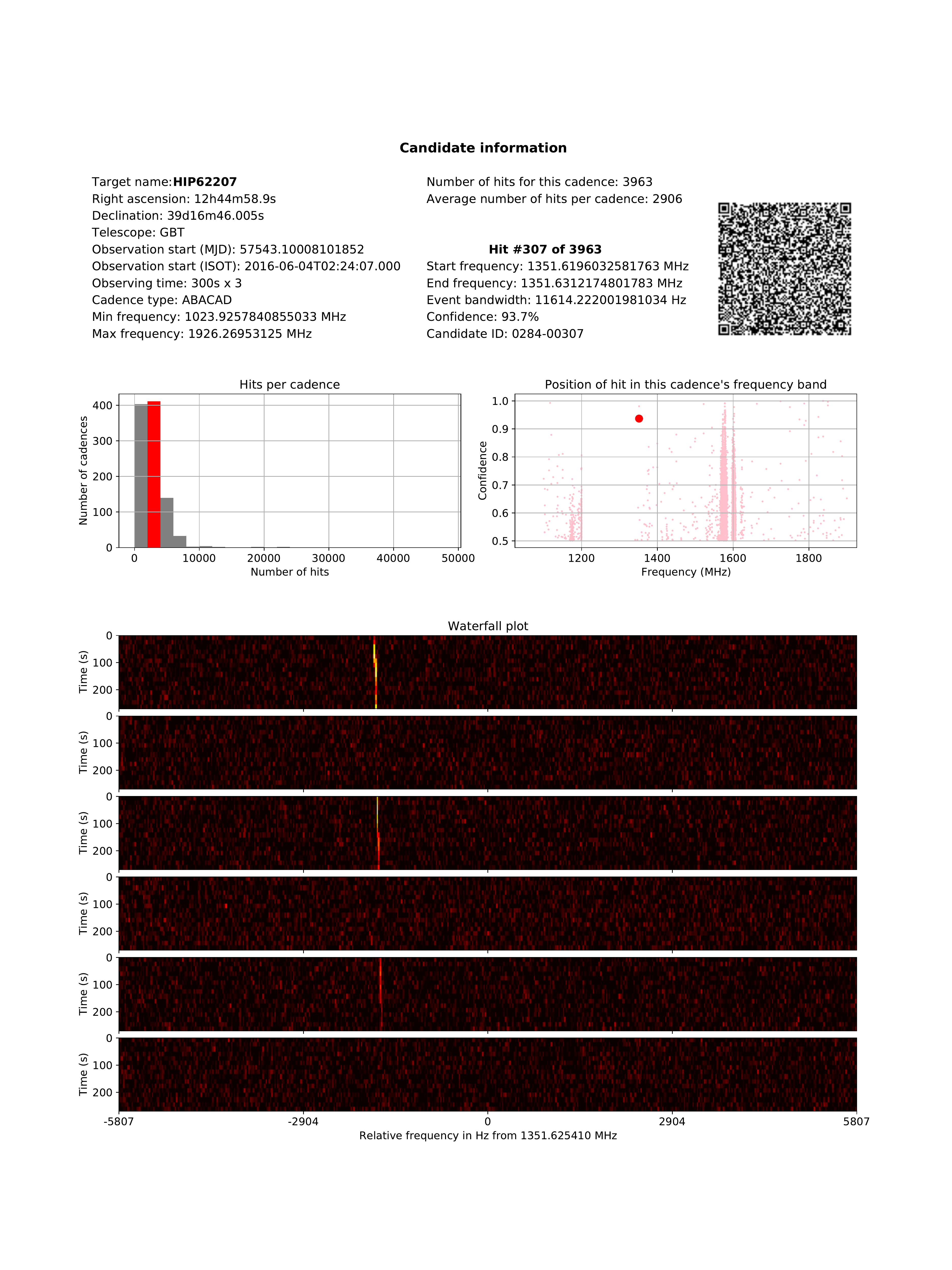}
\end{subfigure}%
\begin{subfigure}{.5\textwidth}
  \centering
  \includegraphics[width=0.95\linewidth]{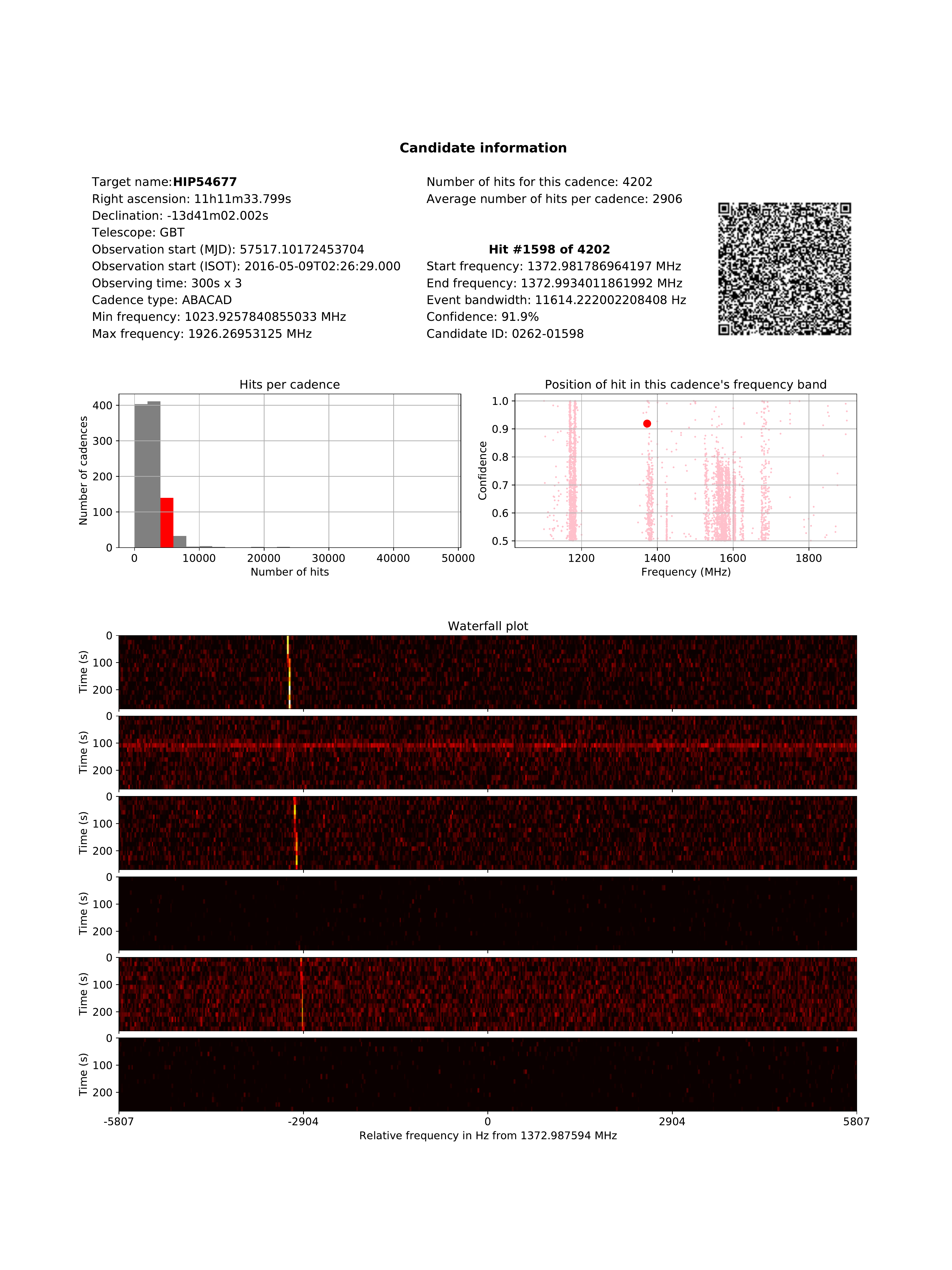}
\end{subfigure}
\caption{Diagnostic plots for (Left) MLc3 (Right) MLc4.}

\label{fig:MLc3-MLc4}
\end{figure}

\begin{figure}[H]
\begin{subfigure}{.5\textwidth}
  \centering
  \includegraphics[width=0.95\linewidth]{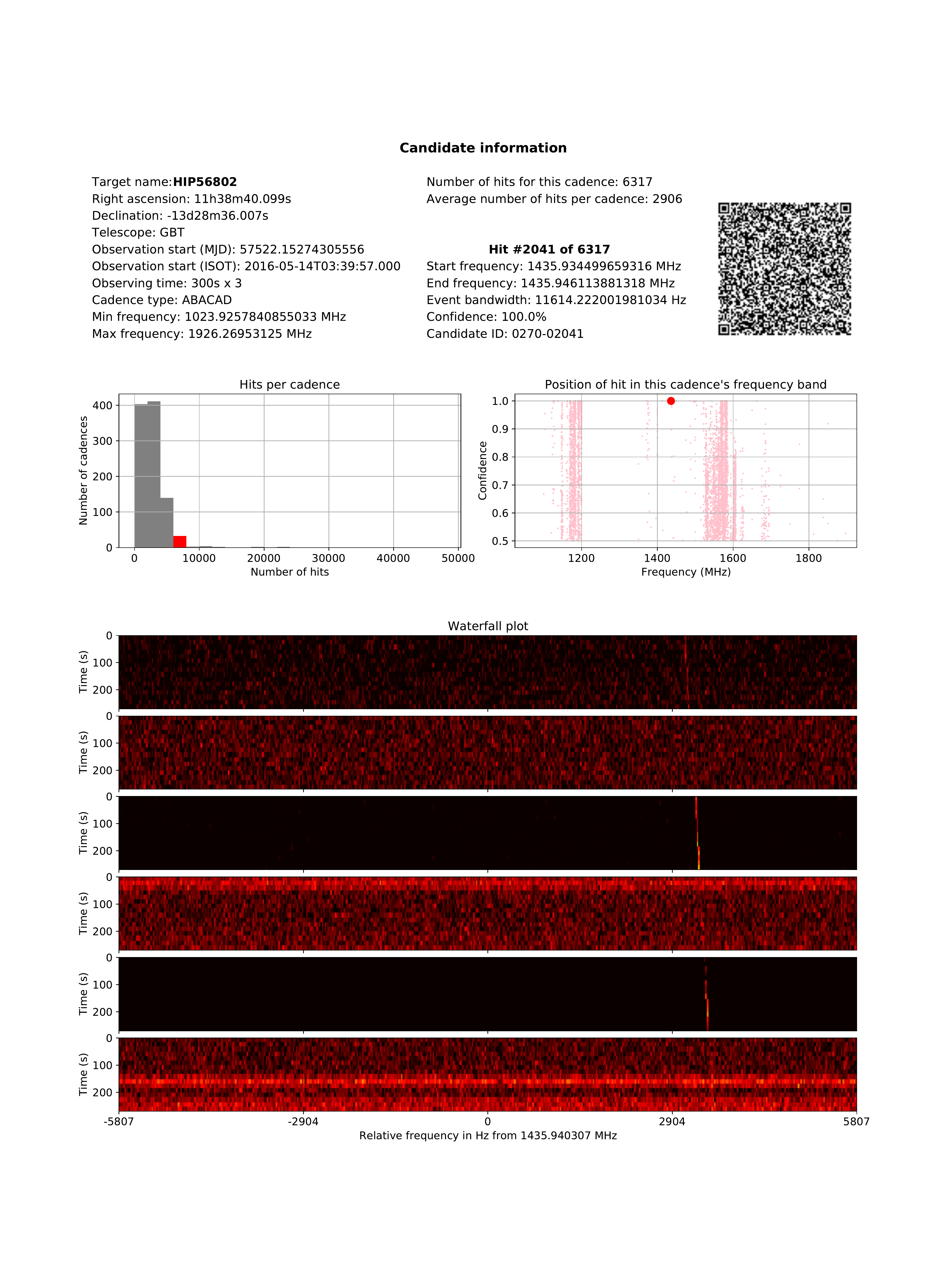}
\end{subfigure}%
\begin{subfigure}{.5\textwidth}
  \centering
  \includegraphics[width=0.95\linewidth]{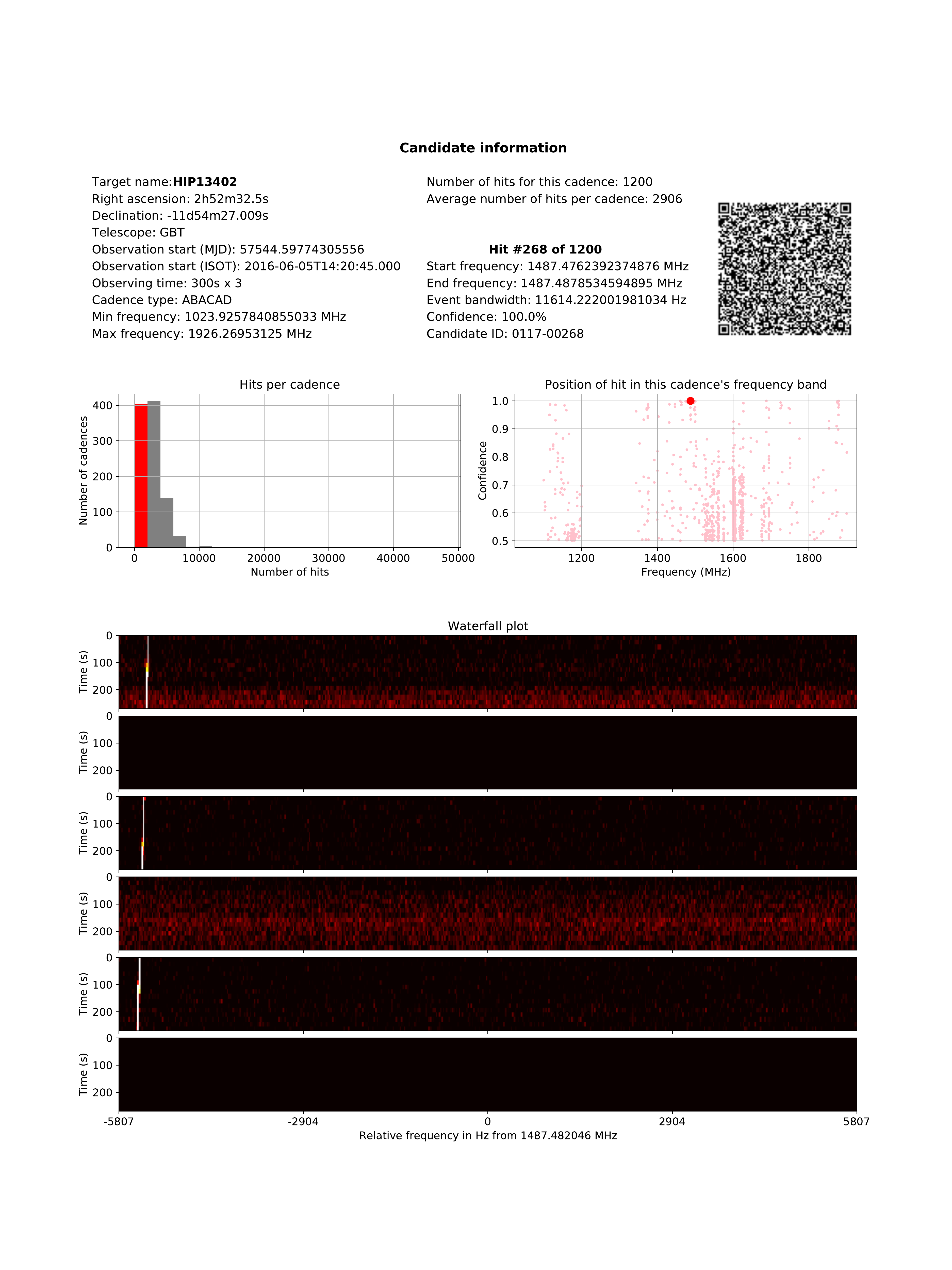}
%  \caption{}
\end{subfigure}
\caption{Diagnostic plots for (Left) MLc6 (Right) MLc7.}
\label{fig:MLc6-MLc7}
\end{figure}

\begin{figure}[H]
  \centering
  \includegraphics[width=0.47\linewidth]{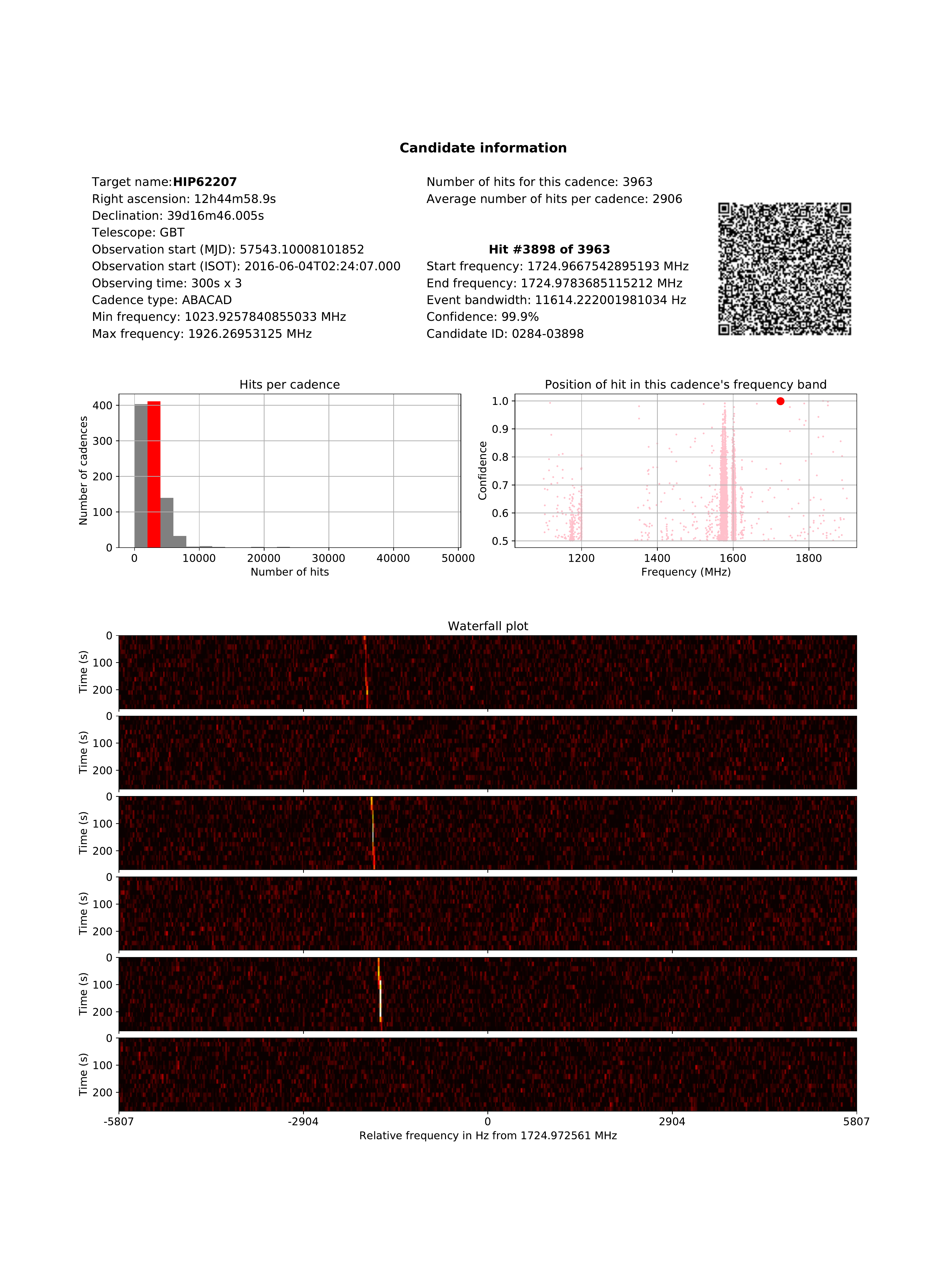}
\caption{Diagnostic plot for MLc8.}
\label{fig:MLc8}
\end{figure}

\end{document}